\documentclass[%
 reprint,
 amsmath,amssymb,
 aps,pra
]{revtex4-2}

\usepackage[utf8]{inputenc}
\usepackage{amsmath}
\usepackage{bm}
\usepackage{graphicx}
\usepackage{mathrsfs}
\newcommand{\Eps}{\mathcal{E}}
\usepackage{hyperref}
\usepackage{color}
\usepackage{tikz}
\usetikzlibrary{shapes.symbols}
\usepackage[export]{adjustbox}
\begin{document}

\title{From Efimov physics to the Bose polaron using Gaussian states}

\author{Arthur Christianen}
\email{arthur.christianen@mpq.mpg.de}
\author{J. Ignacio Cirac}%

\author{Richard Schmidt}%
\email{richard.schmidt@mpq.mpg.de}
\affiliation{%
Max-Planck-Institut für Quantenoptik, Hans-Kopfermann-Str. 1, D-85748 Garching, Germany
}%
\affiliation{%
Munich Center for Quantum Science and Technology (MCQST), Schellingstraße 4, D-80799 Munich, Germany
}%

\date{\today}
\begin{abstract}
Since the Efimov effect was introduced in 1970, a detailed theoretical understanding of Efimov physics has been developed in the few-body context. However, it has proven to be challenging to describe the role Efimov-type correlations play in many-body systems such as quenched or collapsing Bose-Einstein condensates (BECs). To study the impact the Efimov effect can have in such scenarios, we consider a light impurity immersed in a weakly interacting BEC, forming a Bose polaron. In this case, the higher-order correlations are localized around the impurity, making it more feasible to develop a theoretical description. Specifically, we employ a Gaussian state variational Ansatz in the reference frame of the impurity, capable of both capturing the Efimov effect and the formation of the polaron cloud. We find that the Efimov effect leads to a cooperative binding of bosons to the impurity and the formation of a many-body bound state. These many-particle Efimov clusters exist for a wide range of scattering lengths, with binding energies significantly below the polaron energy. As a result, the polaron is not the ground state, but rendered a metastable excited state which can decay into these clusters. While this decay is slow for small interaction strengths, it becomes more prominent as the attractive scattering length increases, up to the point where the polaron becomes completely unstable. This critical scattering length can be interpreted as a many-body shifted Efimov resonance, where the scattering of two excitations of the bath with the polaron can lead to bound state formation. Compared to the few-body case, the resonance is shifted to smaller attractive scattering lengths due to the participation of the polaron cloud in the cooperative binding process. This corresponds to an intriguing scenario of polaron-assisted chemistry \cite{ownpaper}, where many-body effects lead to enhanced signal of the chemical recombination process, which can be directly probed in state-of-the-art experiments.

\end{abstract}

\maketitle

\section{Introduction}

To describe the properties of a many-body system, an accurate understanding of the relevant interactions and correlations between its microscopic constituents is crucial. However, even when the few-body physics is understood, extracting the emergent properties of the system as a whole is not a simple task. A good example is the Efimov effect \cite{efimov:1970,naidon:2017}. This fascinating few-body effect was predicted by Efimov in 1970 in the context of nuclear physics. Efimov showed that there exists an infinite series of three-body bound states appearing close to the unitarity point of a Feshbach resonance with binding energies obeying a geometric scaling law. It took more than 30 years to demonstrate this effect experimentally, first in cold atomic gases \cite{kraemer:2006} and later in Helium \cite{Kunitski:2015}. After that there were more experimental observations and a good theoretical understanding of the few-body physics has been developed \cite{naidon:2017}. However, the task to understand the effect that Efimov-like correlations have on quantum many-body systems has proven to be tremendously challenging and has attracted interest in the context of quenched or collapsing Bose-Einstein condensates (BECs) \cite{makotyn:2014,piatecki:2014,eismann:2016,klauss:2017,colussi:2018,eigen:2018,incao:2018,colussi:2020,musolino:2021}, especially at unitary interactions. Here the rapid buildup of three-body and higher-order correlations renders the description a challenge whose solution remains elusive. 

A different approach to try to understand the effect of Efimov-like correlations in a many-body setting is to study the problem of a mobile impurity immersed in a weakly interacting BEC, leading to the formation of a Bose polaron. Importantly, this problem exhibits the heteronuclear instead of the homonuclear Efimov effect and the relevant three-body processes always involve the impurity. As a result, the most important three-body correlations are localized around the impurity, making the theoretical description more feasible. In particular, this allows the use of variational approaches \cite{levinsen:2015,shchadilova:2016,yoshida:2018}, which is much more difficult when higher-order correlations throughout the BEC have to be accounted for. Also experimentally, probing the Bose polaron can give new insight, due to the available technique of radiofrequency (rf) spectroscopy for its observation \cite{jorgensen:2016,hu:2016,yan:2019}.

\begin{figure*}
    \centering
    \includegraphics[width=0.85\textwidth]{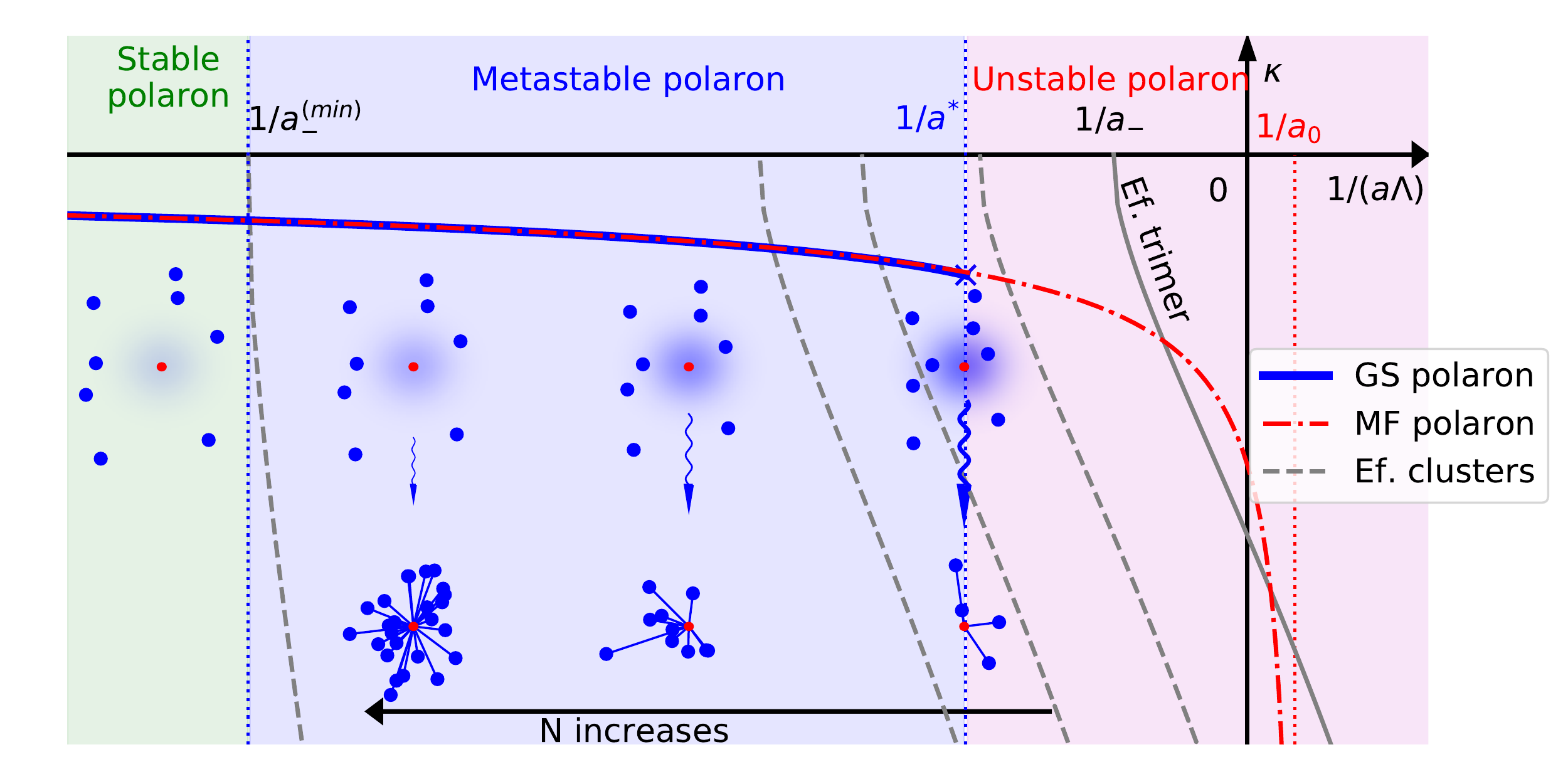}
    \caption{(Color online) Illustration of the main results of this work. In this figure we show the wave number $\kappa=-\frac{\sqrt{M|E|}}{\Lambda}$ of the polaron found using Gaussian States (blue) and mean-field theory (red dash-dotted) as a function of the inverse scattering length $(1/a\Lambda)$. In grey the wave number of the Efimov trimer (solid line), and larger Efimov clusters (dashed lines) is shown. Note that the dashed lines are only schematic and not all possible Efimov clusters are displayed. For small scattering lengths (the green area), the polaron is the ground state of the system and is thus a stable quasiparticle. For $|a|>|a_{-}^{(min)}|$, the smallest scattering length at which an Efimov cluster can be formed, the polaron turns into a metastable excited state, which can decay into the large Efimov clusters via many-body scattering processes. These scattering processes become more likely as the absolute value of the scattering length increases, up to the critical scattering length $a^{\ast}$, at which the polaron is no longer well-defined. At this point scattering of the polaron with one or two more bath particles can lead to decay into an Efimov cluster. In contrast, in the simpler mean-field picture, the polaron remains stable up to the scattering length $a_0$, at which an infinite number of particles pile up in two-body bound states formed with the impurity.}
    \label{fig:introfigPRA}
\end{figure*}

 The concept of the Bose polaron as an impurity immersed in a bath of bosonic excitations was introduced by Landau \cite{landau:1933} to describe an electron dressed by phonons in a solid. A model that is commonly used in condensed matter to describe this scenario is the Fröhlich model \cite{froelich:1954}. However, the Fröhlich Hamiltonian does not contain all necessary terms to describe the cold atom Bose polaron \cite{rath:2013}, since it does not allow for bound state formation. Therefore, to describe processes for which bound state physics is important, such as the Efimov effect, the  Fröhlich model needs to be extended. Furthermore, the description of the wavefunction needs to include the relevant interboson correlations. For this reason, the Bose polaron problem in the cold-atom context has proven to be challenging. Where the Fermi polaron problem can be understood with relatively simple theoretical models \cite{schirotzek:2009,chevy:2006,combescot:2007},  various descriptions of the strong-coupling Bose polaron have yielded widely varying results \cite{levinsen:2015,ardila:2015,shchadilova:2016,grusdt:2017,yoshida:2018,guenther:2021,schmidt:2021}. In Refs.\ \cite{levinsen:2015,yoshida:2018} variational Ansätze with a small number of excitations were used, and a smooth crossover between the polaron and an Efimov cluster was predicted, corroborated by Quantum Monte Carlo results in Ref.\ \cite{ardila:2015}. However, when using a coherent state variational Ansatz which is capable of truly describing a deformation of the BEC instead of only excitations on top of the homogeneous BEC, one finds that the polaron becomes unstable due to an infinite number of excitations piling up on the impurity \cite{shchadilova:2016}. This instability is prevented by explicitly taking into account interboson repulsion \cite{massignan:2006,astrakharchik:2015,chen:2018,guenther:2021,schmidt:2021}, but still the polaron cloud will contain a large number of particles. What the coherent state approach clearly lacks, however, are interboson correlations, meaning the Efimov effect cannot be described. Also, while renormalization group theory \cite{grusdt:2017} predicts the polaron to be unstable already for attractive scattering lengths, no clear connection to Efimov physics is evident.

In this work we systematically extend the coherent state Ansatz by fully including two-body interboson correlations, and three-body correlations including the impurity. To achieve this, we use Gaussian States \cite{shi:2017} in the reference frame of the impurity. As a result, we can both describe the many-body physics of the creation of a macroscopic polaron cloud and the few-body physics of the Efimov effect. In particular, we study how these two phenomena affect each other. We choose to focus on light impurities because in this case the Efimov effect is much more prominent than in the case of equal mass \cite{incao:2006,incao:2006b,tung:2014,pires:2014,naidon:2017,sun:2017}, since a light impurity can more efficiently mediate interactions between the bosons compared to a heavy impurity.

We find that the polaron is no longer the ground state of the extended Fr\"ohlich Hamiltonian when three-body correlations are taken into account, due to the presence of large Efimov clusters much lower in the energy spectrum. However, the polaron remains a metastable excited state up to some critical scattering length $a^{\ast}$, at which the energy barrier protecting the polaron from decaying into an Efimov cluster disappears. This critical scattering length can be interpreted as a many-body shifted Efimov resonance. 

 To explain our results, we first need to understand how the heteronuclear Efimov effect behaves for more than three particles. For the homonuclear case this is well understood \cite{vonstecher:2011,naidon:2017} and four- and five particle Efimov clusters have been experimentally observed \cite{ferlaino:2009,zenesini:2013}. In this case larger and larger clusters are more and more tightly bound. This can be understood from simple dimensional arguments \cite{naidon:2017}. When the atoms have pairwise interactions the binding energy grows with $N^2$, whereas the kinetic energy only grows with $N$. However, the properties of larger Efimov clusters are not as well explored for the heteronuclear case. When the bosons are not interacting or interacting repulsively, and only interact attractively with the impurity, this means that now also the binding energy scales with $N$. This leads to a non-trivial competition between the kinetic and potential energy, and the breakdown of the simple dimensional argument. As a result, providing a general picture of the binding energy of clusters with increasing size, is more difficult for the heteronuclear case. We show here that this effect of larger clusters being more tightly bound still persists for non-interacting bosons when the boson-impurity interaction is modeled with a single-channel model. We will refer to this effect as a ``cooperative binding" effect. This term is imported from chemistry, where it refers to the effect that the particles cooperate with each other to lead to ever stronger binding. This occurs for example in the binding of oxygen molecules to hemoglobin. 

Strikingly, we find that the existence of this cooperative binding effect has a profound impact on the stability of the polaron. We have illustrated this in Fig.\ \ref{fig:introfigPRA}. In this figure we schematically plot the wavenumber (which is proportional to the square root of the energy) of the polaronic state found from Gaussian (blue solid line) and coherent states (red dash-dotted line). The result from coherent states we refer to as mean field theory. The grey solid line indicates the position of the lowest three-body Efimov state in vacuum, appearing at $a_{-}$. The grey dashed lines indicate the lowest Efimov clusters of increasing particle number, which shift to the left as a function of particle number. We prove using simple arguments that there is a minimum scattering length $a_{-}^{(min)}$, needed to form any Efimov cluster which is also indicated in the figure.

The background color of the figure indicates the stability of the polaron. In the green area, no Efimov cluster can be formed yet, meaning that the polaron is the ground state of the extended Fröhlich Hamiltonian. When crossing $a_{-}^{(min)}$ into the blue area, the polaron will no longer be the absolute ground state, but a metastable excited state. For scattering lengths $|a|<|a^{\ast}|$, the decay of the polaron into Efimov clusters, such as indicated with the wiggly vertical arrows, is extremely slow and needs simultaneous scattering of a large number of bosons on the polaron. However, as the scattering length increases, both the number of particles in the polaron cloud increases and the number of particles needed to form a bound state decreases. For both of these reasons the decay processes will become more and more important. When $a^{\ast}$ is crossed into the red area, scattering of one or two additional bosons is sufficient to cause the breakdown of the polaron into an Efimov cluster. Since these decay processes are included in our Gaussian state Ansatz, the local minimum of the polaron on our variational manifold disappears, meaning it is no longer a metastable quasiparticle. Since the BEC provides a coupling between the different particle number clusters, we find a cascade of the wavefunction into ever larger Efimov clusters. 

Remarkably, $|a^{\ast}|$ is smaller than $|a_{-}|$, because the polaron can decay immediately into larger clusters than the Efimov trimer, needing smaller scattering lengths due to the cooperative binding. In experiments, the formation of these large Efimov clusters leads to the rapid chemical recombination into more deeply bound states not included into our model. Furthermore, the shift of the Efimov resonance, means that the experimentally observed resonant recombination signal will also be shifted. This provides a fascinating example of chemistry in a quantum medium and polaron could-enhanced reactivity \cite{ownpaper}.

The energy of the polaron calculated with Gaussian states closely follows the energy of the mean field polaron, up to the scattering length $a^{\ast}$. If no three-body correlations are included, the polaron remains stable even across unitarity up to the scattering length $a_0$. Here the mean field polaron breaks down due to the piling up of an infinite number of bosons in a two-body bound state with the impurity. The Gaussian state polaron breaks down due to decay into an Efimov cluster.

The structure of the main body of our paper is the following: after showing our theoretical methods in section \ref{sec:theory}, we prove in section \ref{sec:coopbind} that our Gaussian state Ansatz incorporates Efimov physics. We present evidence for the cooperative binding effect, extending to large particle numbers, and we elucidate its mechanism. We demonstrate that many-body Efimov clusters form the ground state of the Hamiltonian. Then, in section \ref{sec:polaron}, we use the Gaussian sate Ansatz to calculate the energy of the polaron as a function of scattering length and density and demonstrate its abrupt instability. Finally, we will discuss in detail how our results can be probed experimentally.

\section{Theoretical methods} \label{sec:theory}

\subsection{The Hamiltonian}

To describe the impurity immersed in a three dimensional, infinite, weakly-interacting BEC at zero temperature, we use the  extended Fröhlich model introduced in Ref.\ \cite{rath:2013}.

\begin{multline}\label{eq:original_Hamiltonian}
\hat{\mathcal{H}}_0=\int d\bm{k} \frac{\bm{k}^2}{2m} \hat{a}^\dagger_{\bm{k}} \hat{a}_{\bm{k}}+ \frac{\hat{\bm{P}}^2}{2M}+ g \int d\bm{r} \delta(\bm{r}-\hat{\bm{R}}) \hat{a}^{\dagger}_{\bm{r}} \hat{a}_{\bm{r}} \\ + \frac{g_{B}}{2} \int \int d\bm{r} d\bm{r}' \delta(\bm{r}-\bm{r'}) \hat{a}^{\dagger}_{\bm{r'}} \hat{a}^{\dagger}_{\bm{r}} \hat{a}_{\bm{r'}} \hat{a}_{\bm{r}} .
\end{multline}

Here we treat the impurity with mass $M$ in first quantization, with position and momentum operators $\hat{\bm{R}}$ and $\hat{\bm{P}}$. The bosons of mass $m$ are described using second quantization, where operators $\hat{a}^{\dagger}_{\bm{k}}$ and $\hat{a}_{\bm{k}}$ create/annihilate bosons of momentum $\bm{k}$ that interact via a contact interaction with scattering length $a_B$ determined by a coupling strength $g_B$. The boson-impurity interaction is modeled using a regularized contact interaction and coupling strength $g$, with corresponding scattering length $a$. In Eq.\ (\ref{eq:original_Hamiltonian}), $\int d\bm{k}$ is shorthand for $\int \frac{d^3k}{(2 \pi)^3}$. We have used a single-channel model to describe the interactions. This is supposed to be a good description of the interactions in case in the vicinity of a broad Feshbach resonance.

We model the bosons within the Bogoliubov approximation and introduce the quasiparticle operators $\hat{b}^{\dagger}_{\bm{k}}$ and $\hat{b}_{\bm{k}}$. The validity of this approximation will be discussed in detail in the following sections. After the Bogoliubov rotation, we apply the unitary Lee-Low-Pines transformation \cite{lee:1953} $\hat{U}_{LLP}$ to move to the frame of the impurity,
\begin{equation}
\hat{U}_{LLP}=\exp(-i \hat{\bm{R}} \int d\bm{k} \ \bm{k} \ \hat{b}^\dagger_{\bm{k}} \hat{b}_{\bm{k}}).
\end{equation}

Crucially, this transformation reduces the number of degrees of freedom of the problem, by leading to a transformed Hamiltonian $\hat{\mathcal{H}}=\hat{U}_{LLP}^{\dagger} \hat{\mathcal{H}}_0 \hat{U}_{LLP}$, namely:

\begin{multline}\label{eq:Hamiltonian_Bog}
\hat{\mathcal{H}}=\int d\bm{k} \   \omega_k \hat{b}^\dagger_{\bm{k}} \hat{b}_{\bm{k}} + \frac{(\bm{P}-\int d\bm{k} \ \bm{k} \ \hat{b}^\dagger_{\bm{k}} \hat{b}_{\bm{k}})^2}{2M} \\
  +g n_0+  g \sqrt{n_0} \int d\bm{k} \ W_k [\hat{b}^{\dagger}_{\bm{k}} +\hat{b}_{-\bm{k}}]  \\
+ g \int \int d\bm{k} d\bm{k'} \ \big[ V^{(1)}_{k,k'} \hat{b}^{\dagger}_{\bm{k}} \hat{b}_{\bm{k'}}
+  \frac{V^{(2)}_{k,k'} }{2}  (\hat{b}^{\dagger}_{\bm{k}} \hat{b}^{\dagger}_{\bm{k'}}+\hat{b}_{-\bm{k}} \hat{b}_{-\bm{k'}}) \big],
\end{multline}

where $n_0$ is the density of the BEC. The definitions of the quasiparticle dispersion $\omega_k$, and the variables, $W_k$,$V^{(1)}_{k,k'}$ and $V^{(2)}_{k,k'}$ are given in the appendix \ref{app1}. Since the impurity momentum operator $\hat{\bm{P}}$ now commutes with the Hamiltonian, it has been replaced by a vector $\bm{P}$, which is the total momentum of the system. We set $\bm{P}=0$. Normal ordering the remaining quartic term gives rise to a modified dispersion of the quasiparticles: $\omega_k+\frac{k^2}{2M}$. 

To regularize the boson-impurity contact interaction, we introduce a UV cutoff $\Lambda$ on all momentum integrals, leading to:
\begin{equation} \label{eq:g-renorm}
g^{-1}=\frac{\mu_r}{2\pi a}-\frac{\mu_r \Lambda}{\pi^2},
\end{equation}
where $\mu_r=\frac{m M}{m+M}$ denotes the reduced mass. The value of $\Lambda$ is related to the effective range of the potential \cite{richardthesis}: $\Lambda^{-1}$ is proportional to the Van der Waals length $l_{vdw}$. It therefore also determines the so-called three-body parameter, which is crucial for Efimov physics, since it sets the position of the first Efimov resonance $a_{-}$. The value of $\Lambda$ in our model should be chosen so that the obtained value of $a_{-}$ matches the experimental value of the system of interest.

To reduce computational complexity, we make use of the spherical symmetry of the problem and transform from plane wave modes to spherical waves:
\begin{equation}
\hat{b}^\dagger_{\bm{k}}=(2 \pi)^{3/2} k^{-1} \sum_{lm} i^{l} Y^{\ast}_{lm}(\bm{\Omega}_k) \hat{b}^{\dagger}_{klm}.
\end{equation}

The Hamiltonian then takes the form:
\begin{multline} \label{eq:fullhamiltonian}
\hat{\mathcal{H}}=\sum_{lm}  \int_k (\omega_k+\frac{k^2}{2M})  \hat{b}^{\dagger}_{klm} \hat{b}_{klm} + \\ \int \int d\bm{k_1} d\bm{k_2} \frac{\bm{k}_1 \cdot \bm{k}_2}{2M}   \ \hat{b}^{\dagger}_{\bm{k_1}} \hat{b}^{\dagger}_{\bm{k_2}} \hat{b}_{\bm{k_1}} \hat{b}_{\bm{k_2}} \\
+g n_0 +\frac{g\sqrt{n_0}}{\pi \sqrt{2}} \int_k  k \ W_k (\hat{b}^{\dagger}_{k00}+\hat{b}_{k00}) \\
+ \frac{g}{2 \pi^2}  \int_{k_1} \int_{k_2} \big[k_1 k_2 \  V^{(1)}_{k_1,k_2} \hat{b}^{\dagger}_{k_100} \hat{b}_{k_2 00} \\ +  \frac{V^{(2)}_{k_1,k_2}}{2} (\hat{b}^{\dagger}_{k_100} \hat{b}^{\dagger}_{k_2 00}+\hat{b}_{k_100} \hat{b}_{k_2 00}) \big]
 ,
\end{multline}

where we have defined: $\int_{k}=\int_0^{\Lambda} dk$.

The second term, which we will denote by $\hat{\mathcal{H}}_{QLLP}$,
is written without the transformation to spherical waves. The full expression after the transformation is very lengthy, and given in Appendix \ref{app2}.

This quartic term will later be shown to be extremely important to describe the Efimov effect. Indeed it is the only quartic term of the Hamiltonian, thus it is fully responsible for describing the effective interboson interactions mediated by the impurity. Because this term contains the vector product $\bm{k_1} \cdot \bm{k_2}$, it contains the information about the relative direction of the bosonic momentum, allowing the bosons to distribute kinetic energy more efficiently. This manifests itself in a coupling between consecutive angular momentum terms.

Note that even though we have set $\bm{P}=0$, this does not mean that the impurity is stationary. Even though $\langle \hat{\bm{P}} \rangle =0$, $\langle \hat{\bm{P}}^2 \rangle$ is not zero in the lab frame, and the kinetic energy of the impurity is crucially important for the results.

\subsection{Gaussian states} \label{sec:gausstates}

Throughout this work we use a variational method based on a Gaussian state variational Ansatz. Using Gaussian states to describe the Bose polaron was already suggested in Ref.\ \cite{shchadilova:2016b}, but in their description the beyond-Fr\"ohlich parts of the Hamiltonian were missing. These terms are necessary for describing bound state formation and Efimov physics. Furthermore, the Gaussian part of the wave function was only treated perturbatively. Here we use Gaussian states in the formalism discussed in the review paper by Shi \emph{et al.} \cite{shi:2017}. Note that since we use the Lee-Low-Pines transformation, this entangles the impurity degrees of freedom with bosons, effectively giving a beyond-Gaussian Ansatz in the lab frame. The notation we use is most similar to the presentation in Ref.\ \cite{shi:2019}

The variational Ansatz in the reference frame of the impurity has the form:

\begin{equation}
 |GS\rangle= \exp(\hat{\bm{\Psi}}^{\dagger}\Sigma^z \bm{\Phi}) \exp(i \hat{\bm{\Psi}}^{\dagger} \xi \hat{\bm{\Psi}})  |BEC\rangle.
 \end{equation}
 In this equation $\hat{\Psi}_{klm}=(\hat{b}_{klm},\hat{b}^{\dagger}_{klm})^T$ is the Nambu vector containing the bosonic creation and annihilation operators, $\Sigma^z =\sigma^z \delta(k-k')\delta_{l,l'}\delta_{m,m'}$ with Pauli matrix $\sigma^z$, $\Phi_{klm}=\langle \hat{\Psi} \rangle =(\phi_{klm},\phi^{\ast}_{klm})^T$ is the coherent displacement vector, and $\xi$ is the Hermitian correlation matrix. We do not denote the state of the impurity, because in the frame of the impurity this is trivial: the impurity is stationary at the center of the frame. Furthermore, our Ansatz acts on the state $|BEC\rangle$, which is the state of the background BEC.
 
 The variational parameters we use here are the coherent displacement $\bm{\Phi}$ and the covariance matrix $\Gamma_{kl m,k'l'm'}=\langle \{ \delta\hat{\Psi}_{klm}, \delta\hat{\Psi}^{\dagger}_{k'l m'} \} \rangle$ of the fluctuation field $ \delta\hat{\Psi}_{klm}=\hat{\Psi}_{klm} -\Phi_{klm}$. The covariance matrix $\Gamma$ is related to $\xi$ via the symplectic matrix $S=\exp(i\Sigma^z \xi)$ as $\Gamma=S S^{\dagger}$.  
 We use two complementary algorithms to optimize the variational parameters and to find the minima on our variational manifold: imaginary time evolution and iterated Bogoliubov theory. In principle these two algorithms are equivalent \cite{guaita:2019}, but dependent on the situation either one is computationally more efficient. The equations of motion (EOM) for the imaginary time evolution are given by \cite{shi:2017}:
\begin{align}
    \partial_{\tau}\Phi&=-\Gamma (\eta, \eta^{\ast})^T, \\
    \partial_{\tau} \Gamma &= \Sigma^z \mathscr{H} \Sigma^z - \Gamma \mathscr{H} \Gamma,
\end{align}
 where we define
\begin{align}
    E&=\langle GS|\hat{\mathcal{H}}|GS\rangle, \\
    \eta_{klm}&=\frac{\partial E}{\partial \phi^{\ast}_{klm}}, \\
    \mathscr{H}_{klm,k'l',m'}&=\frac{2\partial E}{\partial \Gamma^{\ast}_{klm,k'l'm'}}. 
\end{align}
The expectation value of the Hamiltonian can be calculated using Wick's theorem.
With iterated Bogoliubov theory, one sequentially solves $\eta=0$ for $\phi$ to update the displacement and diagonalizes $\mathscr{H}$ using a symplectic matrix $S$, to update $\Gamma=S S^{\dagger}$. Performing a single iteration of this algorithm is equivalent to solving the Gross-Pitaevskii equation (solving for $\phi$) and calculating the equivalent of the Lee-Huang-Yang correction (diagonalizing $\mathscr{H}$) \cite{shi:2019}.
This is especially efficient compared to imaginary time evolution in case of very slow dynamics and when $\eta$ is linear in $\phi$. \\

As we will discuss in the following, for some of our calculations we will fix the expectation value of the particle number during the optimization. This works only for the imaginary time evolution, where it is achieved by adding a chemical potential term $\mu_N \hat{N}$ to the Hamiltonian, with particle number operator $\hat{N}$. Then the chemical potential $\mu_N$ is dynamically adjusted to keep the particle number fixed. This can be done using \cite{shi:2020}:
\begin{equation}
    \mu_N=\frac{\langle \hat{N} \hat{\mathcal{H}}\rangle - \langle \hat{N} \rangle \langle \hat{\mathcal{H}}\rangle}{\langle \hat{N}^2 \rangle-\langle \hat{N} \rangle^2}.
\end{equation}

\section{Efimov clusters and cooperative binding} \label{sec:coopbind}

 Cooperative binding is an important aspect of the Efimov effect, since a three-body Efimov state is bound when no two-body bound state can be formed yet. In this section we demonstrate that this cooperative binding can in fact persist for an arbitrary number of particles. We consider the scenario of an impurity interacting with bosons without interactions among themselves via a single-channel model. The cooperative character becomes evident from our finding that adding an extra particle to a bound state also lowers the energy of the other particles, or in other words, that a $N+1$-body bound state is easier to form than an $N$-body bound state. In terms of the illustration in Fig.\ \ref{fig:introfigPRA}, this means that the grey dashed lines move further and further to the left for increasing $N$. These strongly bound Efimov clusters are lower in energy than the polaron and therefore the polaron is not the ground state but an excited state of the Hamiltonian.

\subsection{The two-and three-body problem}
\subsubsection{Variational Ansätze}
Before moving on to the description of large Efimov clusters, it is instructive to first consider the two- and three-body problem (\emph{i.e.} with one, or two bosons plus the impurity). To achieve this we set the background density $n_0$ in Eq.\ (\ref{eq:fullhamiltonian}) to zero, and we consider non-interacting bosons ($a_B=0$). This yields the Hamiltonian

\begin{multline}\label{eq:Hamiltonian_simp}
\hat{\mathcal{H}}=\sum_{lm}  \int_k \frac{k^2}{2\mu_r}  \hat{b}^{\dagger}_{klm} \hat{b}_{klm}  +\hat{\mathcal{H}}_{QLLP} \\
+ \frac{g}{2 \pi^2}  \int_{k_1} \int_{k_2}  k_1 k_2\ \hat{b}^{\dagger}_{k_100} \hat{b}_{k_2 00} .
\end{multline}

For the two-and three-body problem, we use the exact variational Ansätze
\begin{align}
    |\psi_{2b}\rangle& = \int_k \beta_k \hat{b}^{\dagger}_{k00} |0\rangle, \label{eq:two-bodyansatz} \\
    |\psi_{3b}\rangle &=\frac{1}{\sqrt{2}} \sum_{lm} \int_{k_1}\int_{k_2} (-1)^m \alpha_{k_1k_2l} \hat{b}^{\dagger}_{k_1 l m} \hat{b}^{\dagger}_{k_2 l -m} |0\rangle. \label{eq:three-bodyansatz} 
\end{align}
In these expressions angular momentum conservation has been taken into account and the total angular momentum has been set to zero. For the two-body problem (Eq. (\ref{eq:two-bodyansatz})), this means that the single boson always has angular momentum zero, whereas in the three-body problem (Eq. (\ref{eq:three-bodyansatz})), the two bosons always have opposite angular momenta in the frame of the impurity. Moreover the prefactor only depends on $m$ with the sign $(-1)^m$, because of the spherical symmetry.

Despite the complicated form of $\mathcal{H}_{QLLP}$, the EOM for real-time evolution are relatively simple when these symmetries are implemented, and one finds

\begin{align}
    i\partial_t \beta_k &=\frac{k^2 \beta_k}{2\mu_r}+\frac{gk}{2 \pi^2} \int_{k'} k' \beta_k' \label{eq:2body_eom},
    \\
    i \partial_t \alpha_{k_1k_2l}&=\alpha_{k_1k_2l} \frac{k_1^2+k_2^2}{2 \mu_r} + \frac{g \delta_{l,0}}{2 \pi^2} \int_k k (k_{2} \alpha_{k_1 k 0} + k_{1} \alpha_{k k_20}) \notag \\ &-\frac{k_1k_2}{M(2l+1)} [ (l+1)\alpha_{k_1k_2(l+1)} + l \alpha_{k_1k_2(l-1)}] \label{eq:3body_eom}.
\end{align}

\subsubsection{The non-interacting problem}
First, we consider the non-interacting problem ($g=0$). In the non-interacting problem the momenta of each particle are good quantum numbers. As a result, the two-body problem is trivial, since the total momentum is fixed to zero, and the impurity and the boson will thus always have opposite momenta, giving an energy of $\frac{k^2}{2\mu_r}$. The $\hat{\mathcal{H}}_{QLLP}$-term does not play any role here.

In contrast, for the three-body problem, the EOM in the impurity frame (Eq.\ ($\ref{eq:3body_eom}$)) are non-trivial, since the $\hat{\mathcal{H}}_{QLLP}$ term couples the different angular momentum modes. The lowest energy for fixed momenta of the two bosons $k_1$ and $k_2$ and total momentum $\bm{P}=0$ occurs when the two particles move in opposite directions. To describe this, the directions of movement of the bosons need to be correlated, which is achieved by the coupling between the angular momentum modes.
 
The non-interacting EOM of the three-body problem can be mapped to the problem of two coupled harmonic oscillators. This can exactly be solved using a Bogoliubov rotation. In this case an energy is obtained of $\frac{k_1^2+k_2^2}{2m}+\frac{(k_1-k_2)^2}{2M}$, corresponding to an impurity momentum of $k_1-k_2$.  This means that the bosons indeed move in opposite directions. In the corresponding wave function, every angular momentum mode is equally populated. Remarkably, this implies that an infinite number of modes is needed to solve the non-interacting problem in this framework. When $g<0$, as considered below, lower angular momentum modes are favored, since only the zeroth angular momentum mode has an (attractive) coupling with the impurity. In case of bound-state formation, this means that convergence can be achieved with a smaller number of modes.

\begin{table}[tp]
    \centering
    \caption{(Color online) Scattering lengths $a_{-}$ of the appearance of the lowest three-body Efimov states for $^6$Li interacing with different species of bosons, calculated using a reguralized contact interaction, in units of three-body parameter $\Lambda$.}
    \begin{tabular}{c|cccccc}
    \hline
    \hline
     species  &$^{168}$Er &  $^{133}$Cs& $^{87}$Rb & $^{41}$K & $^{23}$Na &$^7$Li  \\
       \hline
        $a_{-}\Lambda$ & -5.20 & -5.69 & -6.91 & -11.1 & -19.0 & -135 \\
    \hline
    \hline
    \end{tabular}
    \label{tab:ac_species}
\end{table}

\subsubsection{Including interactions}

Now we consider the case of a finite, negative $g$. In the two-body problem, a bound state is formed for positive scattering lengths, which vanishes into the scattering continuum at infinite scattering length. However, for the three particles, due to the Efimov effect, a bound state can already be formed for finite negative scattering lengths. Indeed, already at this point cooperative binding is demonstrated, which is driven by the $\hat{\mathcal{H}}_{QLLP}$-term of the Hamiltonian. When $\hat{\mathcal{H}}_{QLLP}$ is not included (leading to the second line of Eq.\ (\ref{eq:3body_eom})), the EOM become separable, and the EOM of the two-body problem are retrieved. This shows that the $\hat{\mathcal{H}}_{QLLP}$-term must be crucial for the Efimov effect. Since this term originates from the kinetic energy of the impurity, this shows that reducing the kinetic energy of the impurity is the driving force of the Efimov effect. 

The full three-body problem can be solved numerically using sparse matrix diagonalization employing linear or, more efficiently, logarithmic grids in $k$. This way, the ground state and low-lying excited Efimov states can be found efficiently.  In principle this method can be generalized to arbitrary interaction potentials between the bosons and the impurity.  In that case the matrix elements of the potential may need to be precomputed numerically, since they typically do not have a straightforward representation in spherical wave momentum space. 
Including boson-boson interactions in this framework leads to complicated expressions for the relevant matrix elements and removes the sparsity of the matrices that represent the equations of motion, greatly increasing numerical complexity. 

In Table \ref{tab:ac_species} we show the Efimov scattering length $a_{-}$ following from our model for $^6$Li-atoms interacting with several types of bosons. We see that $a_{-}$ is smallest for light impurities, meaning that the cooperative binding is the strongest in that case. This can easily be understood from its mechanism: reduction of the kinetic energy of the impurity. Because the kinetic energy of the impurity for a fixed momentum is larger for light impurities this implies that reducing this energy can have a larger impact. This can also be understood directly from the $\hat{\mathcal{H}}_{QLLP}$-term of the Hamiltonian, which scales inversely with the mass of the impurity. Since $a_{-}$ quickly grows as the mass ratio decreases, observing Efimov physics experimentally can most easily be achieved by immersing a light impurity in a bath of heavy bosons. In this paper, we will use the mass ratio of the experimentally available system of a $^6$Li impurity immersed in a BEC of $^{133}$Cs \cite{pires:2014,tung:2014} as an example.

\subsection{Many-body Efimov problem}
To show that the cooperative binding effect of the Efimov effect also persists for larger particle numbers, one needs to demonstrate that the binding energy per boson $E/N$ monotonically increases with $N$.

In order to study $N$-body bound states, the methods used in the previous subsection need to be extended to higher particle numbers. However, this route of exactly solving the $N$-body problem rapidly becomes computationally intractable. Instead, we adopt a different approach, based on a variational wave function that is not a particle number eigenstate, but rather a superposition of states with different particle numbers: a Gaussian state. In this Ansatz pairwise correlations between the bosons are fully taken into account in the frame of the impurity, which translates to three-body correlations between two bosons and the impurity in the lab frame. Instead of fixing the number of particles, we now fix the expectation value of the particle number $\langle \hat{N} \rangle$. We can show that $E/\langle \hat{N} \rangle$ grows monotonically with $\langle \hat{N} \rangle$, which is a strong indication that also $E/N$ grows with N.

Note that the functional form of the correlations in a Gaussian state, is different from the correlation functions in the Jastrow-formalism \cite{jastrow:1955} often used in condensed-matter physics, which was for example also used to describe the polaron in Ref.\ \cite{ardila:2015,drescher:2020}. In our case, the covariance matrix, which characterizes the correlations, is optimized variationally without restrictions on the functional form of the correlations.

In the following we will now highlight our various findings.

\subsubsection{Gaussian states incorporate 2- and 3-body physics} \label{sec:linearizegaus}

First, we demonstrate that a Gaussian state after the Lee-Low-Pines transformation is capable of describing Efimov physics. To this end, we show that the EOM of the two-and three-body problem can be retrieved exactly  by linearizing the EOM for the Gaussian states around the vacuum. The procedure of linearizing the EOM is described in detail in Ref.\ \cite{shi:2019} and we retrace here the main steps. Without loss of generality, $\xi$ can be written as
\begin{equation}
    \xi=\begin{pmatrix}
    0 & \delta \xi \\
    \delta\xi^{\dagger} &0 
    \end{pmatrix}.
\end{equation}

The equations of motion (EOM) for real time evolution are given by \cite{shi:2019}
\begin{align}
    i\partial_{t}\phi&=\eta, \\
    \partial_{t} \Gamma &= \Sigma^z \mathscr{H} \Gamma - \Gamma \mathscr{H} \Sigma^z.
\end{align}

The linearized EOM of $\Gamma$ can be rewritten in terms of $\delta\xi$
\begin{equation}
    i\partial_t \delta\xi=\{D,\delta \xi \}-i (S_0^{\dagger} \delta \mathscr{H} S_0)_{12}.
\end{equation}
In this equation, $D$ is obtained by symplectic diagonalization of $\mathscr{H}(\phi_0,\Gamma_0)$ by a matrix $S_0$ with $\Gamma_0=S_0 S_0^{\dagger}$. The subscript $12$ indicates the off-diagonal block in the Nambu basis.

Writing
\begin{equation}
    \mathscr{H}=\begin{pmatrix}
    \Eps & \Delta \\
    \Delta^{\dagger} &\Eps^{\ast} 
    \end{pmatrix},
\end{equation}

we can diagonalize our equations around the vacuum to find

\begin{equation}
    i\partial_t \delta\xi=-i\Delta+\delta \xi \Eps^{\ast}+\Eps \delta \xi.
\end{equation}
Expressions for $\eta$, $\Eps$ and $\Delta$ are given in the appendix, which are functions of the quantities:
\begin{align}
    G_{k_1 k_2 l}&=\langle \delta\hat{\psi}_{k_2lm}^{\dagger}, \delta\hat{\psi}_{k_1lm} \rangle, \\
    F_{k_1 k_2 l}&=(-1)^m\langle \delta\hat{\psi}_{k_1l-m} \delta\hat{\psi}_{k_2lm} \rangle.
\end{align}.

Note that despite the fact that the right-hand side of these expressions contains the angular momentum label $m$, the matrix elements of $\bm{G}$ and $\bm{F}$ formulated this way do not depend on $m$.

Finally, expanding to first order in $\delta \phi$ and $\delta \xi$ we find:
\begin{align}
    &i\partial_t \delta \phi_k =\frac{k^2 \delta\phi_k}{2\mu_r}+\frac{gk}{2 \pi^2} \int_{k'} k' \delta\phi_k', \\
   & i \partial_t \delta\xi_{k_1k_2l} =\delta\xi_{k_1k_2l} \frac{k_1^2+k_2^2}{2 \mu_r} + \frac{g}{2 \pi^2} \delta_{l,0} k_{2} \int_k k \ \delta\xi_{k_1 k 0} \notag \\ & + \frac{g}{2 \pi^2} \delta_{l,0} k_{1} \int_k k \ \delta\xi_{k k_20} +i \frac{k_1k_2(l+1)}{M(2l+1)} \delta F_{k_1k_2(l+1)} \notag \\
& \quad +i\frac{k_1 k_2 l}{M(2l+1)}\delta F_{k_1k_2(l-1)}.
\end{align}
Since $\delta F= i \delta \xi$ when linearizing around the vacuum, it is evident that these equations exactly correspond to the EOM for the two- and three-body problem: Eq.\ (\ref{eq:2body_eom}) and \ref{eq:3body_eom}.

\subsubsection{Cooperative binding with Gaussian states}
To find evidence for the cooperative binding effect, we optimize the ground state energy of the system by imaginary time evolution. In this procedure we include a dynamically changing chemical potential to fix the average number of particles (not the density), given by:
\begin{equation}
    \langle GS | \hat{N} | GS \rangle = |\bm{\phi}|^2 +\mathrm{Tr}(G).
\end{equation}
 During the optimization with fixed particle number, the coherent part of the wave function goes to zero, leaving a purely Gaussian State. This is not surprising, since this allows for the maximum amount of correlation between the particles. 
 
  \begin{figure}[!t]
     \centering
     \includegraphics[width=0.48 \textwidth]{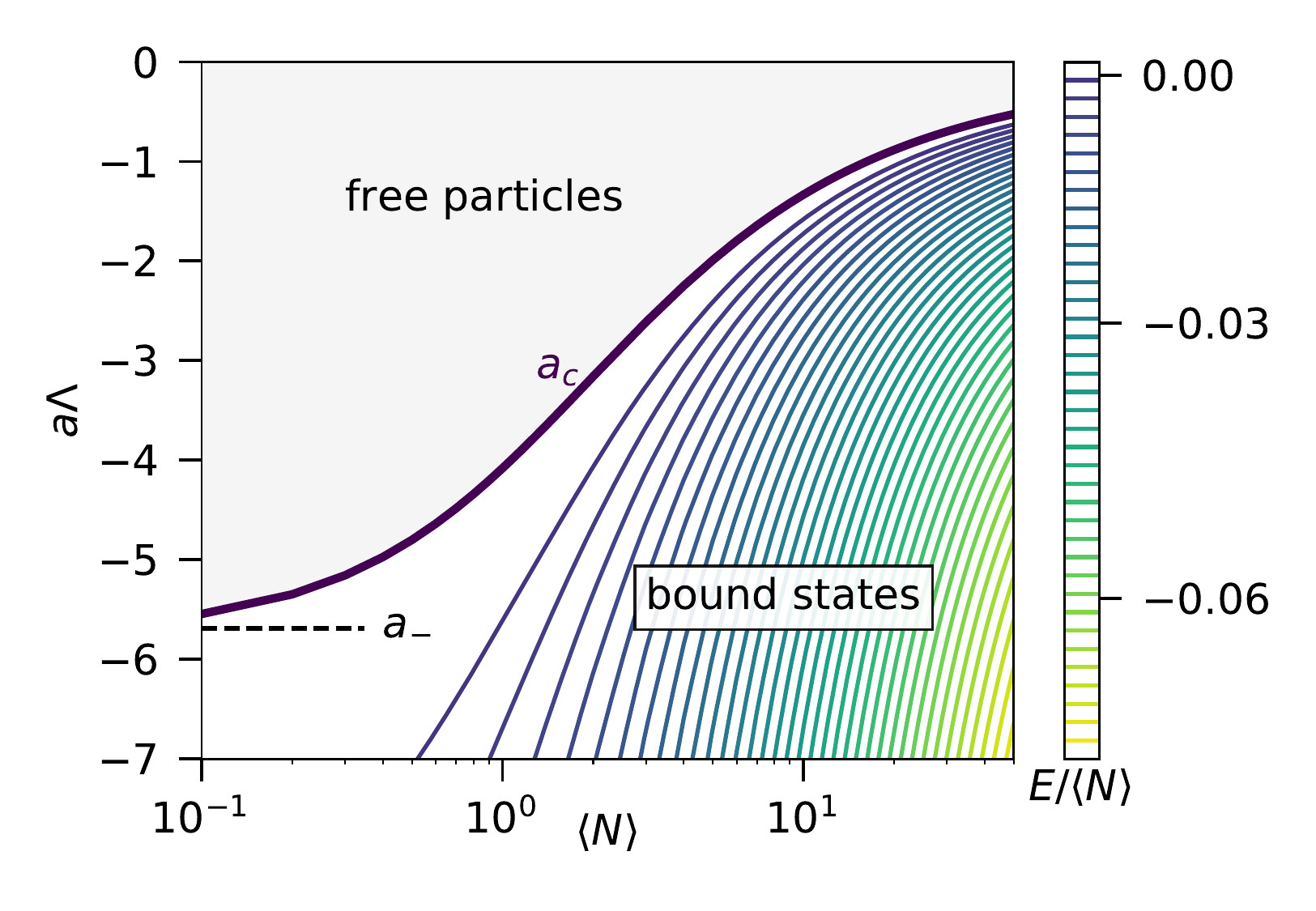}
     \caption{(Color online) The energy per particle $E/\langle N \rangle$ as a contour plot as a function of the number of particles $\langle N \rangle$ and the scattering length $a$ in terms of the three-body parameter $\Lambda$, for $M/m=6/133$. The energy is given in units of $\Lambda^2/M$. The bold line indicates the critical scattering length at which a bound state first appears.}
     \label{fig:coop_binding}
 \end{figure}
 
 In Fig.\ \ref{fig:coop_binding} we show a contour plot of the energy per particle as a function of $\langle \hat{N} \rangle$ and $a$. The bold line indicates the scattering length where the energy per particle becomes negative: the critical scattering length $a_c$ for bound state formation.   Note that since the Hamiltonian conserves particle number,  $a_c$ is a smooth line only due to a classical average over different particle number sectors. In the limit of $\langle \hat{N} \rangle$ going to zero, $a_c \rightarrow a_{-}$, the Efimov scattering length at which a 3-body bound state appears. This is expected, since we showed previously that in the vacuum limit the EOM for the 3-body problem are exactly recovered. As $\langle \hat{N} \rangle$ is increased, $|a_c|$  monotonically decreases. Importantly, as soon as $\langle \hat{N} \rangle$ is sufficient to lead to bound state formation, $E/\langle \hat{N} \rangle$ monotonically increases with $\langle \hat{N} \rangle$ at fixed $a$, clearly demonstrating cooperative binding.  This result seems in good correspondence with recent work by Blume \cite{blume:2019}, who also shows a strongly increasing energy as a function of particle number.

  It is important to realize that our method is variational not for fixed $N$ but for fixed $\langle \hat{N} \rangle$. This means that the energy we find for a given $N=\langle \hat{N} \rangle$ may be lower than the lowest particle number eigenstate of $N$ particles. This is not because of a coupling between the particle number eigenstates in the Hamiltonian, but because $E(\langle \hat{N} \rangle)$ is a concave function. This means that having a larger spread in particle number gives rise to a lower energy. Furthermore, the particle number fluctuations do not become relatively small for large $\langle \hat{N} \rangle$, such as often the case in statistical physics or for coherent states. Indeed, the standard deviation of the particle number for the Gaussian states scales with $\langle \hat{N} \rangle$. This standard deviation is actually maximized, since including a larger spread lowers the energy. The Gaussian form of the wave function does restrict the particle number statistics, meaning not arbitrary superpositions of particle number eigenstates are allowed. These features of our variational results with Gaussian states may be surprising, but do not invalidate our strong evidence for the cooperative binding effect.

 One obvious question is whether $a_c$ will go to zero if $\langle \hat{N} \rangle$ goes to infinity. A hint of how to approach this question is provided by the insight that the cooperative binding effect is driven by the quartic LLP-term originating from the kinetic energy of the impurity. A lower bound on $|a_c|$ can therefore be found by setting the kinetic energy of the impurity to zero. In this case the Hamiltonian becomes:
\begin{equation}\label{eq:Hamiltonian_alim}
\hat{\mathcal{H}}=\sum_{lm}  \int_k \frac{k^2}{2m}  \hat{b}^{\dagger}_{klm} \hat{b}_{klm}
+ \frac{g}{2 \pi^2}  \int_{k_1} \int_{k_2} k_1 k_2 \hat{b}^{\dagger}_{k_100} \hat{b}_{k_2 00} .
\end{equation}
Formally, this corresponds to the Hamiltonian of bosons interacting with a infinitely heavy impurity. There is however an important subtlety: the relation between the scattering length and the coupling constant is determined by the mass of the impurity as described in Eq.\ (\ref{eq:g-renorm}). In our effective model, we therefore get a different scattering length $a_{eff}$ corresponding to an impurity with infinite mass, which is related to $g$ via
\begin{equation} \label{eq:g_lim}
    g^{-1}=\frac{m}{2\pi a_{eff}}-\frac{m \Lambda}{\pi^2}.
\end{equation}
In the spectrum of the Hamiltonian of Eq.\ \ref{eq:Hamiltonian_alim} a bound state crosses the continuum when $a_{eff}=-\infty$. Inserting this into Eq.\ \ref{eq:g_lim} immediately yields the effective scattering length
\begin{equation} \label{eq:aclim}
    a_{c,\mathrm{lim}}=-\frac{\pi M}{2m\Lambda}.
\end{equation}
For Li-Cs this value is given by $a_{c,\mathrm{lim}}=-0.07\Lambda^{-1}$. As can be seen in Fig.\ \ref{fig:coop_binding}, this limit is only approached extremely slowly, and might not be reached with a Gaussian state variational Ansatz. In other words, the scattering length at which the most deeply bound Efimov state can form $a_{-}^{(min)} \leq a_{c,lim} $. We stress that this limit is fundamental and independent of the variational method. With more advanced Ansätze the limit can only be approached faster or more closely.

In any case, from comparison with Table \ref{tab:ac_species} it is evident that $a_{c,\mathrm{lim}}$ is approximately two orders of magnitude smaller than $a_{-}$. Note that such a minimal required scattering length to form an Efimov cluster does not exist for the homonuclear case, since there the pairwise interactions between all of the participating particles drive the cluster formation.

\begin{figure}[tp]
     \includegraphics[width=0.5 \textwidth,trim={0.5cm 0 0 0},clip,left]{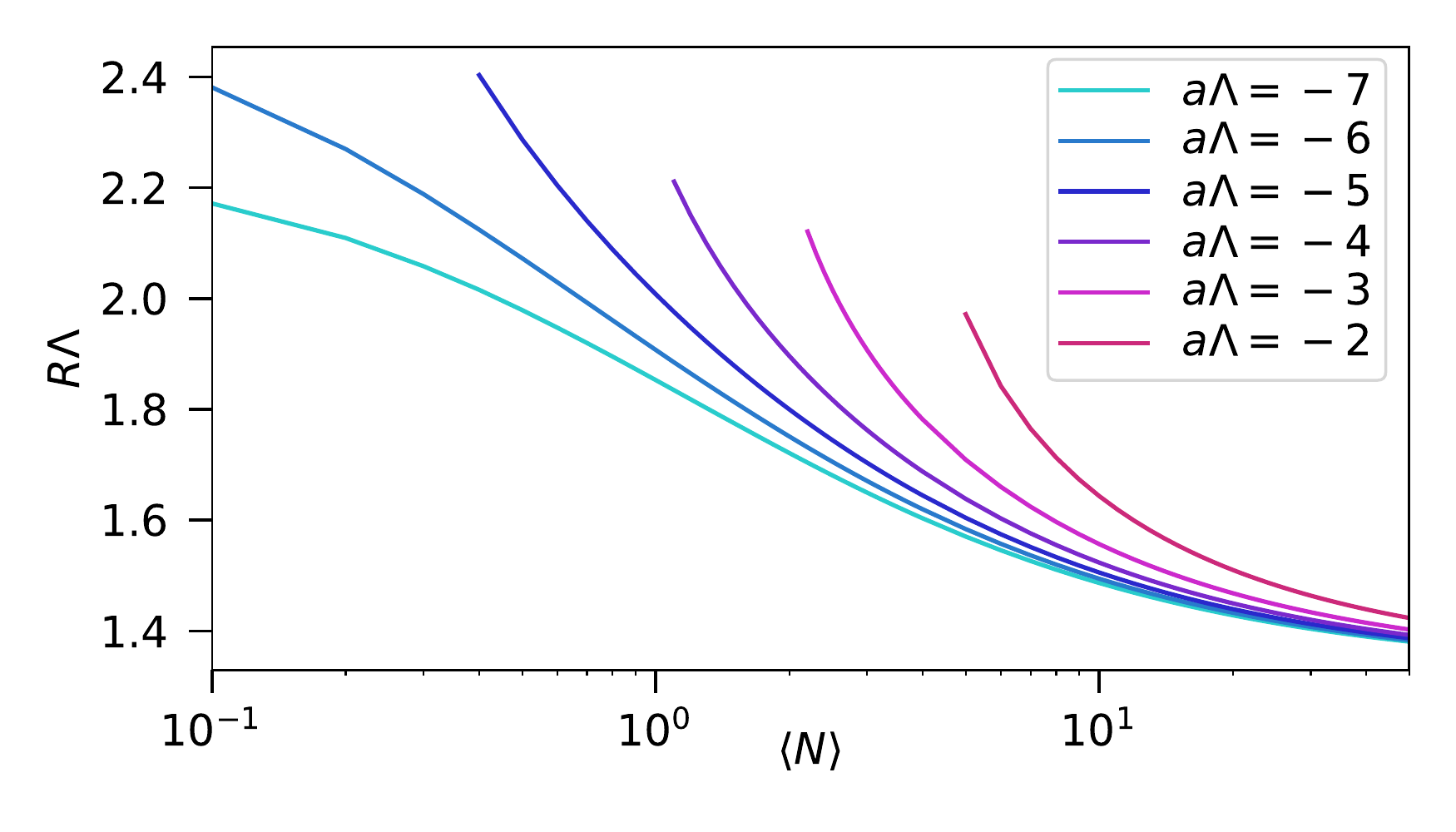}
     \caption{(Color online) Spatial extent  $R$ (defined in Eq.\ (\ref{eq:spatial_extent})) of an Efimov cluster as a function of the number of particles $\langle \hat{N} \rangle$ and the scattering length (indicated by color). The mass ratio $M/m=6/133$.}
     \label{fig:Ef_cluster_size}
 \end{figure}

The increase in binding energy as a function of particle number and scattering length shown in Fig.\ \ref{fig:coop_binding} leads to a decrease in the spatial extent of the Efimov clusters, which we define as
\begin{equation} \label{eq:spatial_extent}
 R=\sqrt{\frac{\langle \hat{N}\rangle}{\langle \int d\bm{k} \bm{k}^2 \hat{b}^{\dagger}_{\bm{k}} \hat{b}_{\bm{k}} \rangle }}.  
\end{equation}
As shown in Fig.\ \ref{fig:Ef_cluster_size},  the size of the Efimov clusters monotonically decreases as a function of $a$ and $\langle \hat{N} \rangle$. Note that $R$ parametrizes the average distance of the bosons to the impurity, and not the distance of the particle furthest away. This means that the total extent of the Efimov cluster wave function can be much larger than shown here, especially close to $a_c$. It is also obvious that these cooperative Efimov clusters are fairly tightly bound, with $R\Lambda$ being of order 1.

\begin{figure}[!t]
     \includegraphics[width=0.47 \textwidth,trim={0.5cm 0 0 0},clip,left]{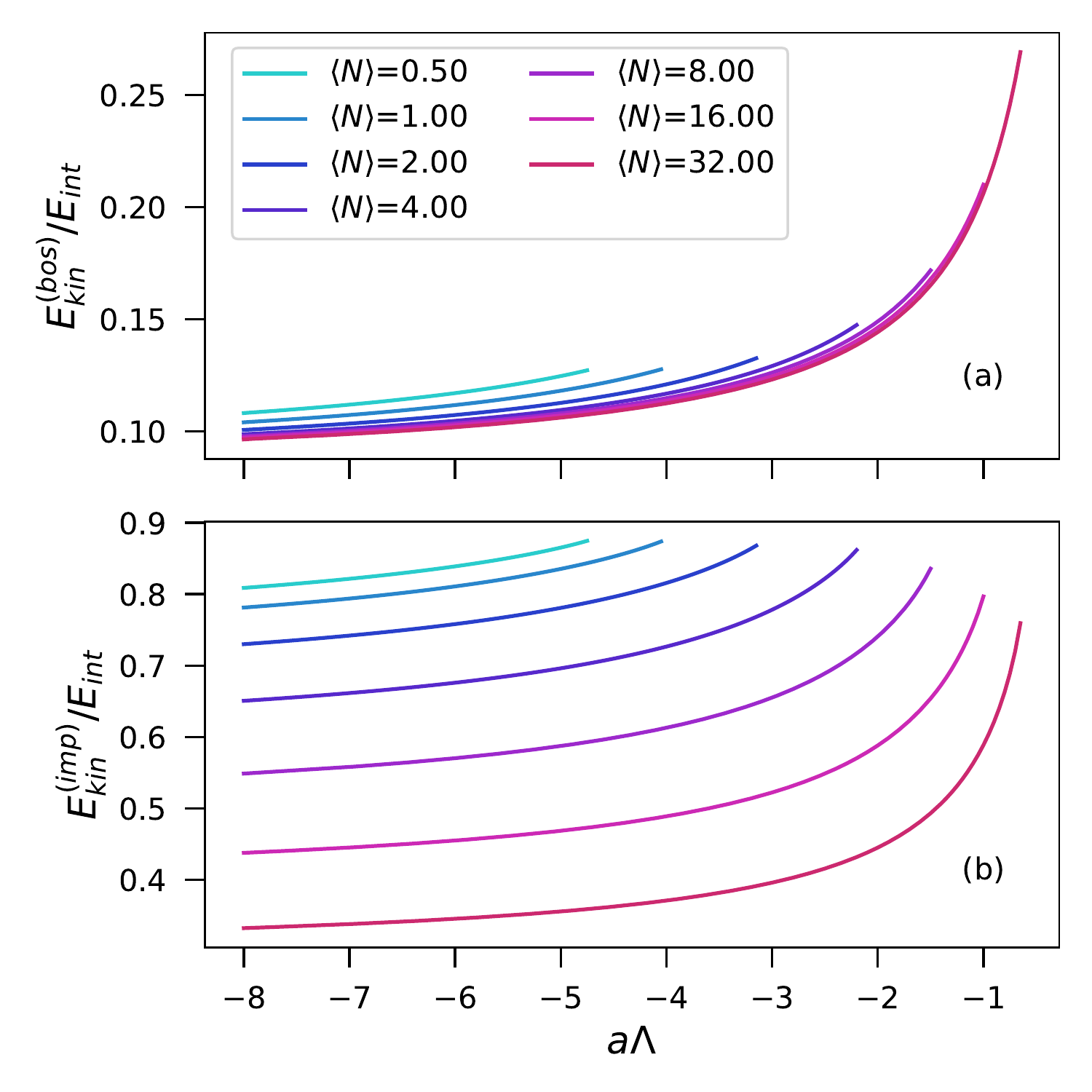}
     \caption{(Color online) The kinetic energy of the bosons (a) and the impurity (b) compared to the total interaction energy, as a function of the scattering length $a$ (in terms of $\Lambda$), for several values of $\langle N \rangle$ and $M/m=6/133$. The legend displayed in figure (a), also corresponds to figure (b).}
     \label{fig:coop_binding_mech}
 \end{figure}

 Finally we corroborate our qualitative explanations of the cooperative binding mechanism with quantitative numerical results. In Fig.\ \ref{fig:coop_binding_mech} we plot the relative values of the different energetic contributions: the kinetic energy of the bosons, the kinetic energy of the impurity, and the interaction energy, as a function of the scattering length and the particle number $\langle N \rangle$.  In this figure we clearly see that the reduction of the kinetic energy of the impurity compared to the interaction energy, is the driving force of the cooperative binding mechanism. Where the ratio between the interaction energy and the bosonic kinetic energy only changes marginally when the number of particles is increased, the kinetic energy of the impurity relative to the interaction energy can decrease with more than a factor two. Note that the kinetic energy of the impurity does not decrease in an absolute sense when more bosons are added. Adding more bosons leads to tighter binding and thus larger kinetic zero-point energy.

\begin{figure*}
\includegraphics[]{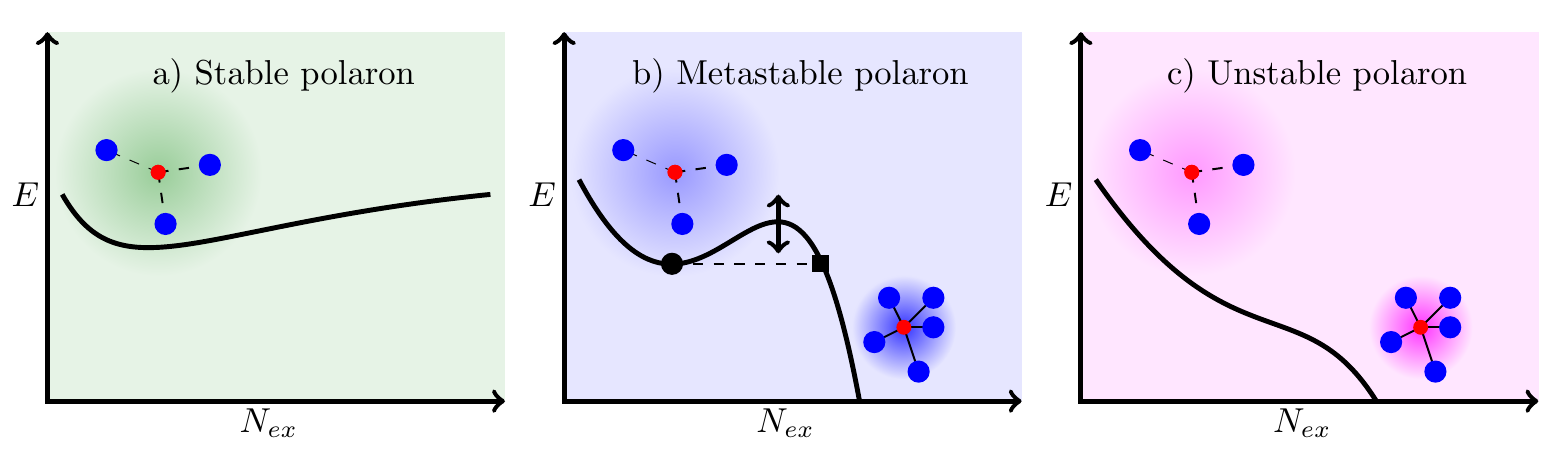}
\caption{(Color online) Illustration of the energy landscape as a function of the number of excitations $N_{ex}$ surrounding the impurity at a fixed density and scattering length in the a) stable, b) metastable and c) unstable regime, which correspond to the regimes shown in Fig.\ \ref{fig:introfigPRA}. In the stable polaron regime, there are no Efimov clusters to decay into, in the metastable regime there is a barrier protecting the polaron from this decay, whereas in the unstable regime this barrier has completely disappeared. The black circle and square in b), connected by the dashed line, are of relevance for Fig.\ \ref{fig:collapse_mechanism}.}
\label{fig:cartoons}
 \end{figure*}

 \subsubsection{Further aspects}
 
 We have not discussed the structure of the many-particle Efimov clusters in detail. One of the reasons is that these clusters are inherently unstable due to recombination into deeply bound molecular states, not included in our model. Because of this rapid recombination, the specific structure of these clusters is not relevant for experiments. Our main interest in this work is to show that the polaron, as discussed in the next section, can decay into these clusters, which is independent of their microscopic structure. For that physics, the cooperative binding effect is essential, making this our main focus.
 
The results we obtain in Figs.\ \ref{fig:coop_binding}, \ref{fig:Ef_cluster_size} and \ref{fig:coop_binding_mech} are not universal since for $N \gg 1$ the bound states that are formed are very tightly bound (as shown in Fig.\ \ref{fig:Ef_cluster_size}). Instead, we predict a breakdown of universality due to the piling up of bosons around the impurity. The cooperative binding mechanism leading to this breakdown is, in turn, universal though, since the driving $\hat{\mathcal{H}}_{QLLP}$-term is independent of the interaction potential.

The accumulation of bosons on the impurity can be prevented or limited by including an interboson repulsion. Due to the tightly-bound nature of the bound states not only the interboson scattering length but also the range of the interboson interactions will play an important role, meaning a simple contact interaction will not be sufficient to describe this effect and more realistic potentials will need to be employed. Whether ultimately Van der Waals universality remains even for deeply bound states and states with more particles, is an interesting open question.

Furthermore, also using a two-channel model, where the interaction between the impurity and the bosons is modeled by a coupling to a closed molecular channel, may modify the results. Especially in the closed-channel dominated limit only a single boson can interact with the impurity at a time, resulting in an effective three-body repulsion \cite{yoshida:2018}. 

Finally note that the coherent part of the Gaussian state may also contribute to cooperative binding, but more weakly than the Gaussian part. This is due to the mixing between the coherent term and the Gaussian term in the expectation value of $\hat{\mathcal{H}}_{QLLP}$. For the next section this is important, because only the coherent state term couples to the linear Fröhlich term in the Hamiltonian, responsible for the polaron formation.

\section{The Efimov effect and the Bose polaron} \label{sec:polaron}

We have shown that large and strongly bound Efimov clusters already form at small negative scattering lengths. The ground state in our system is therefore not the Bose polaron, but a state where all particles are bound to the impurity. Questions concerning the fate of the polaron immediately arise. Does a polaron state still exist and if yes, will it still be stable? Can it decay into the cooperative Efimov clusters? Finding answers to these questions is the topic of this section.

 \subsection{Stable and metastable polaron}
 
 We are now in the position to discuss in more detail the spectrum shown in Fig.\ \ref{fig:introfigPRA}. Starting at the left of this figure, as discussed below Eq.\ (\ref{eq:aclim}) in the previous section, there is a minimum scattering length $a_{-}^{(min)}\leq a_{c,lim}$ needed to form any Efimov cluster.
As a result, for $a>a_{-}^{(min)}$ the polaron in fact remains the ground state of the Hamiltonian. However, $a_{-}^{(min)}$ can be very small for light impurities, yielding only a small region where the polaron is truly stable. 

The first clusters that can form at $a_{-}^{(min)}$ contain an exceedingly large number of particles. Since smaller clusters are not stable yet, this implies that all particles required for the bound state formation need to come together without the possibility to cascade through long-lived intermediate states. Since cold atomic gases are dilute, the rate of the required $N$-body scattering processes is extremely small. Therefore the polaron state, which  only leads to a relatively small and long-ranged deformation of the condensate at small scattering lengths, will not experience rapid decay into these clusters and remain metastable slightly beyond $a_{-}^{(min)}$.
 
 One can also explain mathematically from our Hamiltonian why the polaron should remain a metastable state. For small scattering lengths and a small number of excitations forming the polaron cloud, the contribution of the $\hat{\mathcal{H}}_{QLLP}$ term of the Hamiltonian is negligible. Since we are looking at attractive scattering lengths, no two-body bound states can be formed yet. This implies that in Eq.\ (\ref{eq:Hamiltonian_Bog}) the quadratic kinetic term dominates over the quadratic interaction terms. The model that therefore remains resembles closely the original Fröhlich model with a linear term driving polaron formation and a quadratic term counteracting this. This model must have a minimum of the energy as a function of particle number. This is illustrated graphically in Fig.\ \ref{fig:cartoons}, where the energy landscape of the Hamiltonian is plotted as a function of the number of excitations around the impurity for fixed density. The three subfigures correspond from left to right to the stable, metastable and unstable regime. 
 
 For $0>a>a_{-}^{(min)}$, the polaron is the global minimum of our energy landscape and we are in the regime illustrated in Fig. \ref{fig:cartoons}a). When $a_{-}^{(min)}$ is crossed this does not immediately affect the local minimum since the barrier protecting it is still very large. Only for a very large number of excitations a bound state can be found. Mathematically, as long as the particle number at the local minimum is small enough that the quartic term does not play a big role yet, the assumptions discussed before remain valid. This changes, however, for large scattering lengths or densities. In this case the particle number at the polaron minimum found from this simple model, is so large that the quartic term can no longer be neglected. When this occurs, the barrier protecting the polaron from decaying into Efimov clusters starts to disappear, as depicted in \ref{fig:cartoons}b). Eventually, at critical scattering length $a^{\ast}$, this barrier completely disappears leading to a breakdown of the polaron, entering the regime of \ref{fig:cartoons}c).

\subsection{The breakdown of the polaron}

Now we will discuss in more detail the process leading to the breakdown of the polaron. At the point where the polaronic minimum disappears, the polaron state is no longer stable against a specific type of perturbations. These perturbations can be found from our model by linearizing the EOM such as explained in section \ref{sec:linearizegaus}. The resulting processes then represent single and double excitations on top of the Gaussian State. Thus, the polaron becomes unstable when a single or double excitation (or a linear combination thereof) can result in the formation of a bound state. Once a bound state is formed, the cooperative binding effect directly implies that at that scattering length also larger bound states can be formed with an ever increasing binding energy. This means that more and more bosons will accumulate on the impurity, leading to the eventual collapse of the system onto the impurity.

The value of the scattering length  $a^{\ast}$ where the polaronic instability occurs, can be calculated numerically as a function of the density. To this end, we use the following procedure: we start at small scattering lengths where the local minimum corresponding to the polaron can be easily found. We then incrementally increase the coupling strength. In every step the polaronic properties are calculated with iterated Bogoliubov theory, using the polaron wave function at the previous scattering length as a starting point. As the critical scattering length $a^{\ast}$ is reached, the barrier preventing the polaron from decaying into many-body bound states disappears, leading to an abrupt divergence of the energy and particle number. 

  \begin{figure}[t]
     \centering
     \includegraphics[width=0.48 \textwidth]{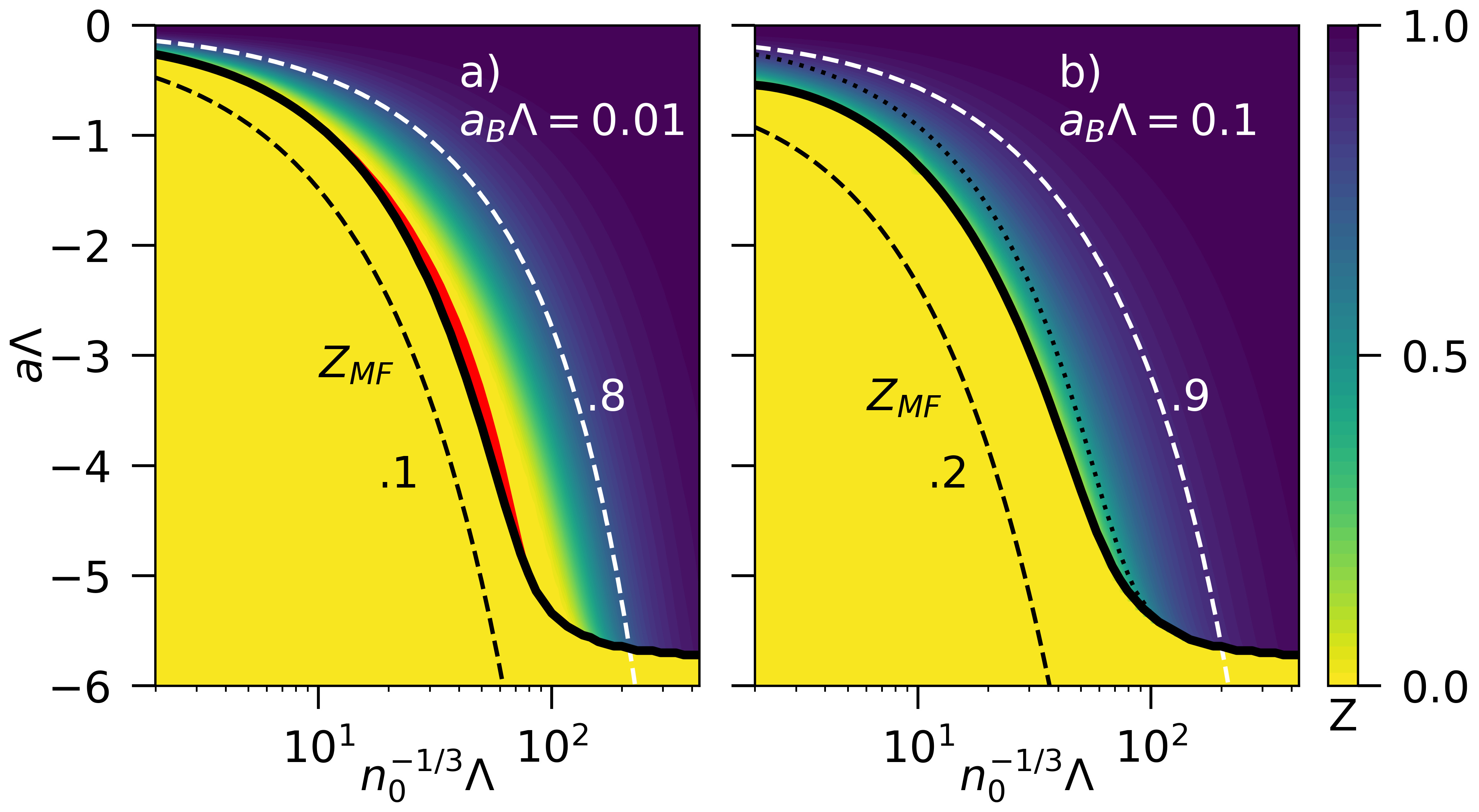}
     \caption{(Color online) The quasiparticle weight $Z$ as a function of the interparticle distance $n_0^{-1/3} \Lambda$ and scattering length $a$ for mass ratio $M/m=6/133$. Two different interboson scattering lengths are chosen: a) $a_B \Lambda=0.01$ and b) $a_B \Lambda=0.1$. The black solid line indicates the critical scattering length $a^{\ast}$ at which the polaron breaks down. The dotted line in b) is just present to more clearly indicate the difference between figure a) and b), and corresponds to the solid line in a). The red region close to the solid line in figure a) is a region of dynamic instability. The dashed lines indicate contours of the quasiparticle weight $Z_{MF}$ obtained from mean-field theory. }
     \label{fig:polaron_stabilityZ}
 \end{figure}
 
 \subsubsection{Polaronic instability and shift of the Efimov resonance}

In Fig.\ \ref{fig:polaron_stabilityZ} we use solid lines to plot $a^{\ast}$ as a function of the interparticle distance $n_0^{-1/3} \Lambda$ for two values of the interboson scattering length $a_B \Lambda=0.01 $ and $a_B \Lambda=0.1$. As a colormap, we additionally show the spectral weight $Z$ of the polaron. The weight $Z$ is given by the overlap of the wave function with the vacuum state, and it can experimentally be measured using injection rf-spectroscopy \cite{jorgensen:2016,hu:2016}. 

For Gaussian states, $Z$ is given by:
\begin{equation}
    Z=\frac{\exp(-\Phi^{\dagger} (\Gamma+I)^{-1} \Phi)}{\sqrt{\det(\frac{\Gamma+I}{2})}},
\end{equation}
where $I$ is the identity matrix.
The red region close to the solid line in Fig.\ \ref{fig:polaron_stabilityZ}a), defines a region of dynamical instability, to be discussed further below.

First we consider the low-density limit, where we find that $a^{\ast} \rightarrow a_{-}$, $\emph{i.e.}$, the polaron ceases to exist exactly at the three-body Efimov scattering length. This finding can be understood as follows. For small densities, the polaron cloud is extremely dilute, meaning the impurity is practically free. A bound state for a free impurity plus two excitations from the background can be formed when the first three-body bound state crosses the continuum: at $a_{-}$. This transition from a free impurity to a three-body Efimov state gives rise to a sharp drop of the quasiparticle weight from 1 to 0.

As the density is increased, a polaron cloud surrounding the impurity forms which contains excitations from the BEC. There are multiple consequences of this polaron cloud formation. First, as can be seen from \ref{fig:polaron_stabilityZ}, it leads to a reduction of the quasiparticle weight for increasing density. Second, due to the increased density around the impurity, scattering on the polaron can immediately lead to the formation of clusters of more than three particles. Importantly, due to the cooperative binding effect these larger clusters can already be formed at smaller scattering lengths than $a_{-}$. This means that, as shown in Fig.\ \ref{fig:polaron_stabilityZ}, $|a^{\ast}|$ shifts to smaller scattering lengths as a function of density. This gives rise to the interpretation of the polaronic instability as a many-body shifted Efimov resonance, where the polaron now takes over the role of the free impurity as a collision partner of two \emph{additional} bosons. This reasoning also implies that the time-scale associated with this instability is on the same order as the time-scale associated with three-body recombination.  

One fascinating aspect of this finding, is that the shift of $a^{\ast}$ is continuous. While the average over particle number sectors in our discussion of cooperative binding in section \ref{sec:coopbind} and Fig.\ \ref{fig:coop_binding} was purely classical, here the Hamiltonian coherently couples the different particle number sectors. This means that instead of obtaining a classical average over three-, four- and five-body Efimov states, the clusters that are formed in a BEC contain a quantum mechanical superposition of these different particle states. This originates from the quantum mechanical nature of the BEC, giving rise to the linear Fr\"ohlich term in the Hamiltonian. This highlights an intriguing aspect of chemistry in a medium of a quantum nature.

 \subsubsection{Mechanism of the polaronic instability}
 
We now study the mechanism leading to the polaronic instability in more detail. To interpret our results it is important to introduce the relevant length scales of the problem.  These are:
\begin{itemize}
    \item the size of the Efimov trimer, which is on the order of $a_{-}$.
    \item the average interparticle distance, parametrized by $n_0^{-1/3}$.
    \item the size of the polaron cloud, which is determined by the modified healing length of the BEC: $\xi_B=(8\pi n_0 a_B \mu_r/m)^{-1/2}$ \cite{guenther:2021}.
\end{itemize}

  \begin{figure}
    \centering
    \includegraphics[width=0.48\textwidth]{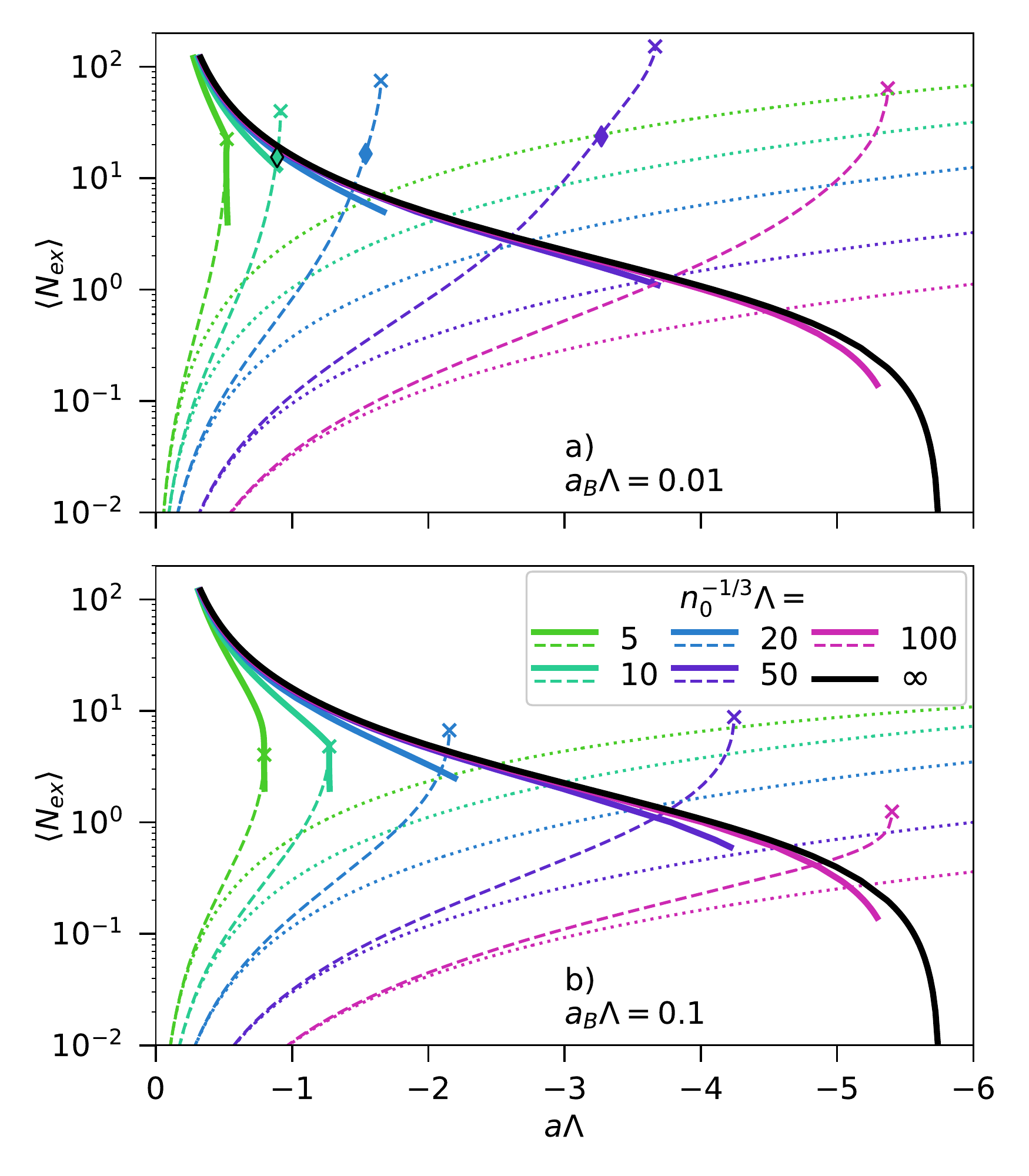}
    \caption{(Color online) The number of excitations $\langle \hat{N}_{ex} \rangle$ contained in the polaron cloud compared to the critical number of particles needed for bound state formation. Specifically, the dashed/dotted lines indicate the number of excitations in the polaron cloud as a function of scattering length $a\Lambda$ using a Gaussian/coherent state variational Ansatz for several background densities. The interboson scattering length is given by a) $a_B \Lambda =0.01$  and  b) $a_B \Lambda =0.1$. The black solid line indicates the critical particle number for bound state formation as a function of the scattering length, corresponding exactly to the line of $a_c$ displayed in Fig. \ref{fig:coop_binding}. The colored solid lines indicate the number of particles needed to form a bound state lower in energy than the polaron as a function of the scattering length in presence of a background condensate. The crosses indicate the points where the polaron is destabilized in the Gaussian state Ansatz. The diamonds indicate the position beyond which a dynamical instability occurs. The legend in figure b) also corresponds to figure a). }
    \label{fig:collapse_mechanism}
\end{figure}


In Fig. \ref{fig:collapse_mechanism} we plot the number of excitations in the polaron cloud computed with a Gaussian (coherent) state Ansatz as a function of the scattering length as dashed (dotted) lines. Along with this, we show as solid lines the critical particle number needed to form a bound state with an energy lower than the polaron energy at the given scattering length and background density. In other words, we compare the number of particles contained in the polaronic excitation cloud with the number of particles that is needed to form a bound state. This figure is closely related to Fig.\ \ref{fig:cartoons}: the dashed and solid lines indicate the position of respectively the black circle and the black square in  Fig.\ \ref{fig:cartoons}b).

The crosses at the end of the dashed lines indicate the scattering length $a^{\ast}$ of the polaronic instability. This instability is not captured by the coherent state Ansatz. In between the crosses and the diamonds that appear on some lines, we find regions of dynamical instability, discussed in more detail below. The black line in this figure shows the critical scattering length for bound state formation as a function of particle number in absence of a background BEC, such as shown in Fig.\ \ref{fig:coop_binding}.  The color indicates the background density. The difference between figure a) and b) is the interboson scattering length $a_B$.
 
 
 First we compare the dotted lines and the dashed lines. We see that the dotted and dashed lines for the same density coincide for small $a$ and $\langle \hat{N}_{ex} \rangle$. However, for larger $a$, the particle number of the polaron calculated using Gaussian states grows much faster. Close to the critical scattering length this difference grows to 1-2 orders of magnitude.
 
 Now we also consider the solid lines. Naively one would expect that the onset of bound state formation would occur when the dashed lines cross the solid lines, but this appears not to be the case. 
 In the low-density regime, the dashed lines in fact cross the solid lines, meaning that the polaron is stable, even when containing more than enough particles to form a bound state. The crucial insight to understand this behaviour, is that $\xi_B \gg a_{-}$. Thus, even though the particle number is very large, the polaron cloud is so extended that the number of particles within the range $a_{-}$ is still too small to facilitate bound state formation. In contrast, in the high-density regime, the instability exactly occurs when the dashed lines hit the solid lines. In this regime the healing length that determines the size of the polaron cloud is smaller than, or of similar size as, the Efimov state. Therefore a bound state can be formed immediately when the polaron cloud contains the required number of particles. In the intermediate density regime, there is a crossover between these two regimes. Consistent with this interpretation, we find that the larger $a_B$, and thus the smaller $\xi_B$, the lower the density at which this transition occurs.
 
We see that $|a^{\ast}|$, indicated with the crosses in  Fig.\ \ref{fig:coop_binding},  shifts towards smaller values for larger densities. From these graphs we find that the main mechanism of this shift is an increase of the number of excitations in the polaron cloud and their density as a function of background density (horizontal shift of dashed lines). Another contribution comes from the vertical shift of the solid lines. This is due to linear Fröhlich interaction term in the Hamiltonian, which leads to stabilization of the coherent polaron cloud due to the background density. This coherent part of the wavefunction also participates in the cooperative binding, leading to a downward shift of the solid lines. This is slightly counteracted by the interbosonic repulsion and the modified quasiparticle dispersion, which plays the strongest role for large densities and $a_B$.

In our framework the most important role of $a_B$ is to determine the healing length. If we compare Figs. a) and b) we see that the healing length mainly determines the number of particles in the polaron. This is also reflected in the quasiparticle weight $Z$ in Fig.\ \ref{fig:polaron_stabilityZ} a) and b).


\subsection{Further aspects}

\subsubsection{Properties of the metastable polaron}

\begin{figure}
     \centering
     \includegraphics[width=0.48 \textwidth]{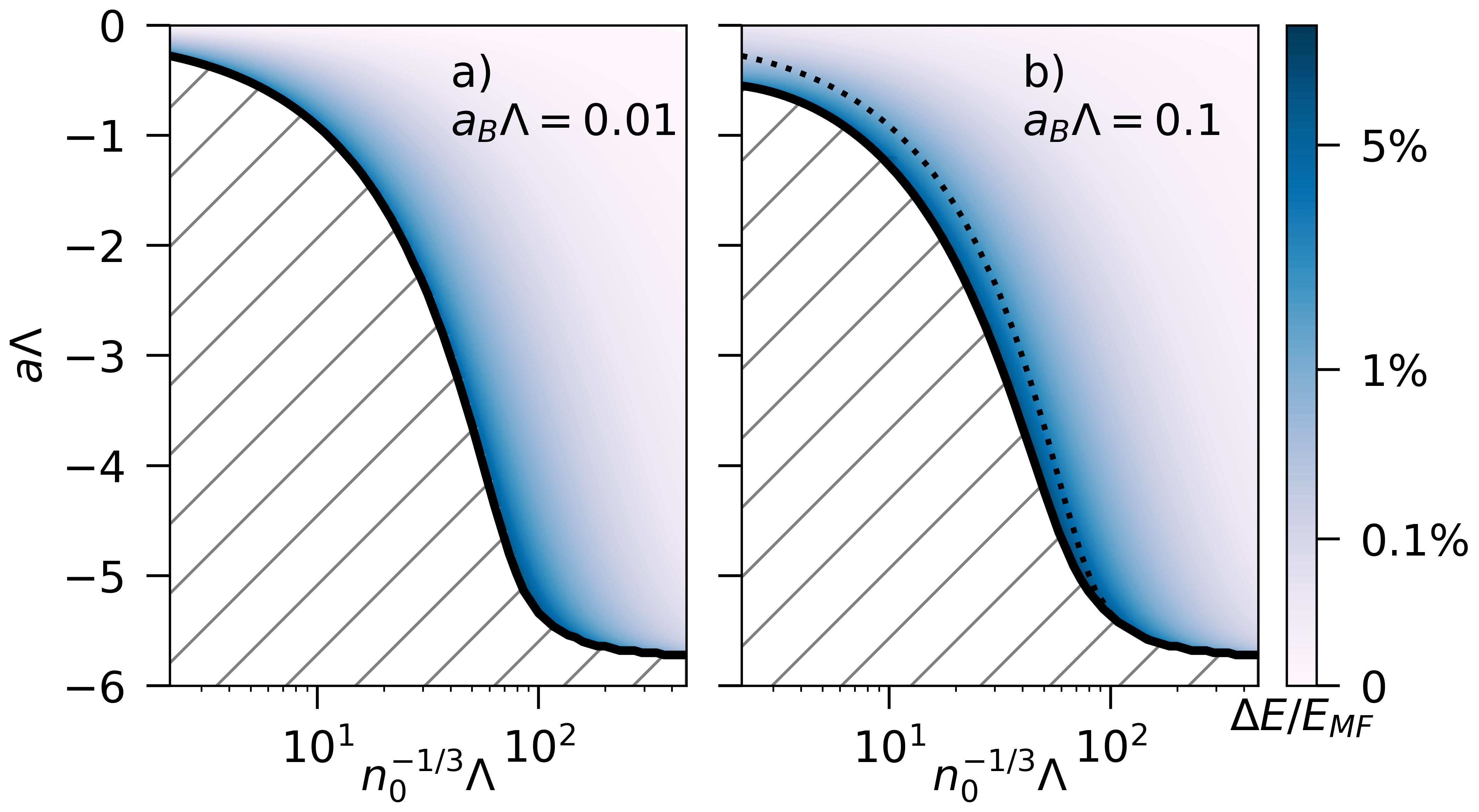}
     \caption{(Color online) The energy difference $(E-E_{MF})/E_{MF}$ as a function of the density $n_0$ and scattering length $a$ for mass ratio $M/m=6/133$ and two different interboson scattering lengths a) $a_B \Lambda=0.01$ and b) $a_B \Lambda=0.1$. The black solid lines indicate $a^{\ast}$. The dotted line in figure b) is shown to more clearly indicate the difference between figure a) and b), and corresponds to the solid line of figure a). Note that the color scale is not linear, but cubic. }
     \label{fig:polaron_stabilityE}
 \end{figure}

Above we have focused on the breakdown of the polaron described by Gaussian states. Another interesting question is how much the properties of the polaron are altered compared to the coherent state approach \cite{shchadilova:2016} by including interboson correlations in the regime where the polaron is metastable. In Fig.\ \ref{fig:polaron_stabilityE} we plot the energy difference $\Delta E=E-E_{MF}$  with the mean field result:
\begin{equation}
E_{MF}=\frac{2\pi n_0}{\mu_r (a^{-1}-a_0^{-1})},
\end{equation}
where $a_0$ is defined in appendix\ \ref{app1}. Surprisingly, the effect on the energy appears to be very small. We see that there is only a significant correction to the polaron energy close to the instability. Even there, the correction is smaller than 10 \%. This leads to the conclusion that introducing correlations in the variational Ansatz leads to a decrease in spectral weight and destabilization at some point, but that the energy of the polaron is still very well described using mean-field theory. 

We have already seen by comparing the dashed and dotted lines in Fig.\ \ref{fig:collapse_mechanism} that the number of excitations in the polaron cloud found from coherent or Gaussian states are very different. For small particle numbers and scattering lengths, they still coincide, since the quartic $\hat{\mathcal{H}}_{QLLP}$-term has only a small contribution. When this term starts to be important, for scattering lengths approaching $a^{\ast}$, the difference rapidly grows however. This also leads to a much smaller $Z$-factor for the Gaussian State result, as can be understood from Fig.\ \ref{fig:polaron_stabilityZ}, by comparing the dashed contours of $Z_{MF}$ compared to the colormap. This reflects the general notion that while variational energies may work well, the same may not apply for the wave functions.

 \subsubsection{Dynamical Instability}
 
Our numerical results show a region of dynamical instability. A dynamical instability occurs when the variational parameters correspond to a minimum on the variational manifold, but to a saddle point of its tangent space. We identify the presence of the dynamical instability by linearizing the real time EOM around the minimum found by the iterated Bogoliubov theory. The system is stable when the symplectic diagonalization of the linearized time-evolution operator yields only positive real eigenvalues. However, in case of a dynamical instability, one finds imaginary eigenvalues, which corresponds to a negative direction in the Hessian. This manifests itself as an instability in the real-time evolution. The dynamical instability indicates that the variational manifold is no longer suitable to describe this metastable state of the system in this region. To solve this problem, one would need to incorporate even higher order correlations in the calculation.
In our Figs.\ \ref{fig:polaron_stabilityZ} the dynamical instability occurs in the small red region attached to the solid line at intermediate densities for $a_B \Lambda=0.01$. This is exactly the region where the polaron is very large in number of particles and extent, but where the density at the impurity is too small to lead to bound state formation. 
 \\
 A dynamical instability can also be found when extending our plots to higher density. Here our results are no longer valid, because when $n_0^{-1/3} \Lambda>1$, one can no longer speak of universal long range physics, and short range physics will dominate the behaviour of the system. This corresponds to interparticle distances comparable in scale to the length scale of the interaction potentials. In gases of cold atoms the typical densities are orders of magnitude away from this limit.
 \\
 A dynamic instability was also found for the Bose polaron with coherent states in Ref.\ \cite{drescher:2018}. Although the character and position of this instability is different from the one we find here, the origin may be related.

 \subsubsection{The dependence on the mass of the impurity}
 
\begin{figure}
     \centering
     \includegraphics[width=0.48 \textwidth]{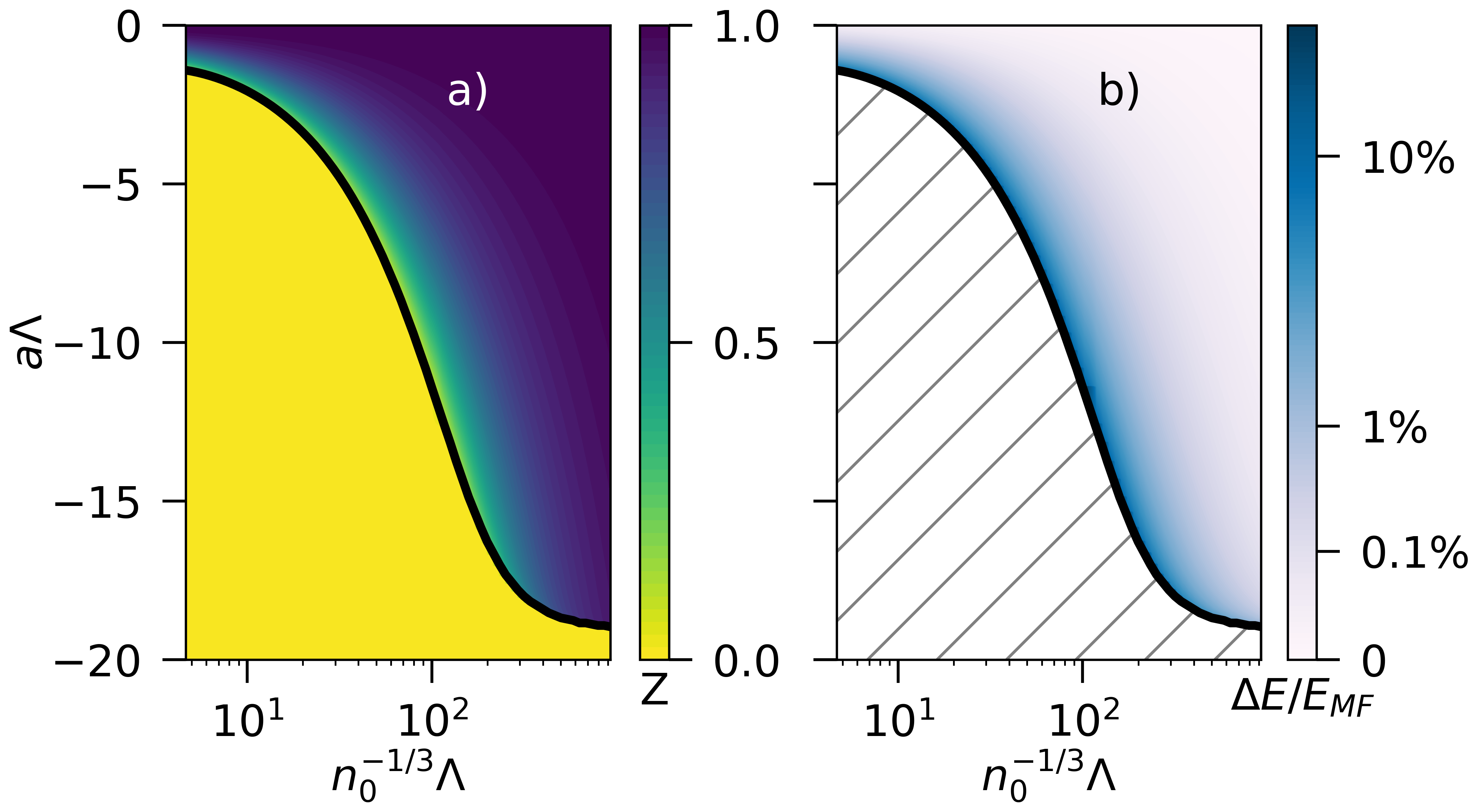}
     \caption{(Color online) a) Quasiparticle weight $Z$ and b) fractional energy difference $(E-E_{MF})/E_{MF}$ as a function of the density $n_0$ and scattering length $a$ for mass ratio $M/m=6/23$ and  $a_B \Lambda=0.1$. Solid lines indicate the critical scattering length $a^{\ast}$. }
     \label{fig:polaron_stability_Z_diffm}
 \end{figure}

 Finally, we study the mass-dependence of our results. To this end, we investigate a significantly different mass ratio and choose the Li-Na system as an example. In this case we fix the interboson scattering length to $a_B\Lambda=0.1$ and we plot $a^{\ast}$ as a function of the density. The result is shown in figure \ref{fig:polaron_stability_Z_diffm}. We plot $a^{\ast}$ both alongside the quasiparticle weight and the energy correction to the mean-field result.

 The most obvious difference is the change in the scale of the scattering length. From the few-body problem we know that $a_{-}$ becomes much larger when mass ratio $m/M$ decreases (see Table \ref{tab:ac_species}) and we find that the same holds in the many-body case. This also leads to a change in the scale of the density, which is not a surprise, since the number which matters is the dimensionless ratio between the interparticle distance and the size of the Efimov state, which scales with $a_{-}$. Except for these changes in scale there are no striking physical differences between the Li-Na and Li-Cs systems studied here. One small quantitative difference is the difference in the polaron energy between the Gaussian state and the mean field theory results. This difference is slightly larger for the Li-Na mass ratio than for Li-Cs.
 
 \section{Discussion and experimental proposal}
 
 \subsection{Limitations of our approach}
 
 Our current approach has some limitations. First, we have only studied very small interboson scattering lengths. In realistic experiments $a_B \Lambda$ will be of order unity. We have limited ourselves to small $a_B$ to preserve the validity of the Bogoliubov approximation. To ensure this, we tested that going beyond the Bogoliubov approximation with these interboson scattering lengths only leads to small deviations from our results ($<10\%$ for $a^{\ast}$). 
 
 Going beyond the Bogoliubov approximation comes with the challenge of strongly increasing numerical complexity and the necessity and difficulty to properly regularize the interboson interactions. Studies dealing with these issues are left for future work. Nevertheless, we expect that for experimentally realistic $a_B$ and for light impurities our results still hold qualitatively. For heavy impurities, however, we expect that the cooperative binding will be strongly suppressed due to the interboson repulsion.
 
 Second, since we have used imaginary time evolution, we could only study the properties of local energy minima on our variational manifold. For that reason, we could in particular not go beyond $a^{\ast}$ to study, $\emph{e.g.}$, higher-order Efimov states or the repulsive polaron. Furthermore, calculating spectral functions with a Gaussian state variatonal manifold is challenging for various reasons. Therefore the fate of the polaron branch in our model beyond the instability remains an important open direction of study. We consider it unlikely that the branch will completely disappear and we hypothesize that it will be broadened due to the decay into the Efimov clusters. We expect that the resulting width of the spectral line will be inversely proportional to the timescale of the decay. At this point this timescale would be similar to the time scale of three-body recombination.
 
Third, our approach considers up to three-body correlations between two bosons and the impurity. It is thus a natural question to ask how our results would generalize when including even higher order correlations. We expect that when up to $N$-body correlations are included, the value of $a^{\ast}$ would asymptotically connect to the resonance of the $N$-body Efimov cluster. Due to the cooperative binding effect, its absolute value would in turn be smaller than the $|a^{\ast}|$ we find here. As discussed before, however, we do not expect the polaron branch in the spectrum to suddenly end at the point of instability, but we expect it to be broadened by the time scale associated to its decay. Since cold atomic gases are typically very dilute, $N$-body scattering in experiments is typically not very prominent for $N>3$. As a result, we expect that the inclusion of beyond three-body correlations would in practice not have a strong effect on the observed results.
 
Finally our results may not fully apply to closed-channel dominated Feshbach resonances, because we use a single channel model. In the case of closed-channel dominated resonances there are effective repulsive three-body interactions counteracting the cooperative binding, which are not included in our model.

 \subsection{Comparison to other theoretical approaches}
 
 We have already compared our results quantitatively with the mean field result obtained from a coherent state approach \cite{shchadilova:2016,guenther:2021}. The most drastic difference is that the coherent state approach does not include three-body correlations, which are needed to describe the Efimov effect, and that it therefore completely misses the cooperative binding effect and the presence of the Efimov clusters. In this model, the polaron would be the ground state for all negative scattering lengths and even beyond unitarity. The polaron energy will then diverge at the positive scattering length $a_0$, due to a large number of excitations piling up on the impurity. The scattering length $a_0$ can be interpreted as a shifted version of the unitarity point, so the point where a two-body bound state crosses the continuum. In contrast to the Gaussian state model, where $a^{\ast}$ is shifted to smaller coupling strengths compared to $a_{-}$, the shift from unitarity to $a_0$ is a shift to stronger coupling strengths. This highlights also the different mechanism of the shifts. For Gaussian states the shift is predominantly caused by the cooperative binding mechanism, whereas the interboson repulsion on the Bogoliubov level causes the shift in case of coherent states.
 
 Interboson interactions play an important role in our model by setting the healing length of the BEC, and hence the extent of the polaron cloud. Guenther \emph{et al.} \cite{guenther:2021} show that explicitly including interboson interactions can prevent the collapse of an infinite number of bosons onto the impurity. However, this only plays a role for large numbers of bosons and high densities in the polaron cloud. In principle, explicit interboson repulsion will also limit the number of particles that an Efimov cluster can host. However, for the light impurities and small scattering lengths we consider, the size of the Efimov cluster where this effect will play an important role is much larger than the Efimov clusters into which the polaron state initially decays. Therefore this interboson repulsion would not qualitatively change the critical scattering length at which the polaron instability occurs. 
 
  Using a renormalization group approach \cite{grusdt:2017}, it was found that the polaron breaks down for a finite negative scattering length.  This was attributed to phase and particle number fluctuations and no connection to the Efimov effect was made.  Since the general picture of a polaronic instability at negative scattering lengths matches with our results, it remains an interesting open question how the two pictures are related.
 
 In Refs.\ \cite{levinsen:2015,yoshida:2018,sun:2017} the interplay between Efimov physics and Bose polaron formation was studied for a limited number of excitations. For a mass ratio $m/M=1$ a smooth crossover between the Bose polaron and the lowest energy Efimov cluster was predicted \cite{levinsen:2015,yoshida:2018} and this was supported by quantum Monte Carlo results \cite{ardila:2015}. In this case the most deeply bound Efimov cluster contained a limited number of particles. In our case the lowest energy Efimov cluster contains an infinite particle number, and there can hence be no smooth crossover from the polaron into this state. These qualitative differences can be explained by the different mass ratios that were used and the effective three-body repulsion which is included in the two-channel model used in Refs.\ \cite{levinsen:2015,yoshida:2018}. In Ref.\ \cite{sun:2017}, radiofrequency injection spectra were predicted for the same mass ratio we considered. However, since the number of possible excitations from the BEC in the employed diagrammatic approach was limited, the possibility of forming many-particle Efimov clusters was not included.

 \subsection{Experimental implementation}
 
 Our predictions have not yet been tested experimentally \cite{hu:2016,jorgensen:2016,yan:2019}. In the Bose polaron experiments that were performed so far, the mass ratios that were used led to a strong suppression of the Efimov effect. As a result, any Efimov features would only have been observable at very large negative or positive scattering lengths. In this work we focus on the example of $^6$Li impurities in a BEC of $^{133}$Cs, but for Li in Rb or K, as also available in experiments \cite{maier:2015, huang:2019}, the results would be very similar. 
 
 The typical densities of above BECs are in the range of $10^{13}$-$10^{15}$ cm$^{-3}$ \cite{pethick:2008}. When we assume that the three-body parameter  $\Lambda \approx l_{vdw}^{-1}$ and we take the Van der Waals-length of Li-Cs of $45 a_0$ this gives a regime of  $n_0^{-1/3} \Lambda \sim 40-200$. This means that the low and intermediate density regime of our results can be probed, and a clear shift of the Efimov resonance should be observable.  Tuning the density in the experiment for a given BEC may in practice be difficult. However, in any case the resonance position in a BEC can be compared with the resonance position in a thermal gas, which should give comparable results to our low density limit.
 
 In the particular case of the Efimov resonances observed for Li-Cs at positive Cs-Cs scattering length \cite{ulmanis:2016,johansen:2017}, there is a subtlety which plays a role. In this case the resonance corresponding to the lowest energy Efimov is suppressed due to coupling to a shallow Cs$_2$ bound state. Therefore the first observed Efimov resonance corresponds to the second Efimov state and appears at a scattering length of around $-2000 a_0$. If we use this Efimov resonance as the lowest resonance in our model, we find a larger three-body parameter  $\Lambda^{-1}=8\ l_{vdw}$. Due to the larger size of the Efimov state, this means that now the high density regime of our results could be probed: $n_0^{-1/3} \Lambda \sim 5-25$. Furthermore, when the mass ratio $m/M$ is smaller such for $^6$Li in a BEC of $^{23}$Na,  the value of $a_{-}$ is naturally larger. This implies that $n_0^{-1/3} a_{-}$ is larger and relatively higher densities are reached as well. 
 
 We propose combining two experimental approaches to test our theoretical predictions. The first approach would be to perform loss measurements such as typically used to observe Efimov resonances \cite{kraemer:2006,ferlaino:2009,zenesini:2013,pires:2014,tung:2014,ulmanis:2016,johansen:2017}. In this case the magnetic field should adiabatically be ramped to form a polaron. Then at a given final scattering length $a$, the magnetic field should be kept fixed to measure the loss arising from recombination. When now this final scattering length is varied, one should observe a clear enhancement of the loss as $a^{\ast}$ is reached. Whether this appears as a resonant feature or as the onset of a regime where the three-body recombination is enhanced, is still and open question and a subject of further study. This may also depend on the density. The loss measurements may be more efficiently performed by using the recently introduced photoassociative ionization technique \cite{eisele:2020}. 
 
 The second approach would be to perform rf-injection spectroscopy such as in Refs.\ \cite{hu:2016,jorgensen:2016} for lighter impurities. We expect that a clear drop in quasiparticle weight and broadening of the spectral line of the polaron will be observed when the scattering length of the polaronic breakdown is approached and crossed. When doing ejection spectroscopy as in Ref.\ \cite{yan:2019} and the ground state polaron is actually prepared, enhanced three-body loss due to the Efimov cluster formation will be the most important observable. The formation of tightly bound Efimov clusters may also give rise to higher frequency contributions in the rf-spectrum because of their large amount of kinetic energy. 
 
 In conclusion, observation of our theoretically predicted phenomena are in reach with current state-of-the-art experimental techniques.
\\

 \section{Conclusion and Outlook}
 
 We use a Gaussian state variational Ansatz to describe the Bose polaron problem and the Efimov effect. We find that the cooperative binding caused by the Efimov effect leads to the formation of many-particle Efimov clusters. The cooperative binding is driven by the reduction of the kinetic energy of the impurity. Since these Efimov clusters are lower in energy than the Bose polaron, the polaron is not the ground state of the extended Fröhlich Hamiltonian but rather exists as a metastable excited state. This excited state loses its stability at a critical scattering length $a^{\ast}$ that can be interpreted as a many-body shifted Efimov resonance.
 We predict that while the mean-field energy of the polaron is reliable up to the point where the polaron becomes unstable, the inclusion of interboson correlations leads to a strong decrease in the spectral weight.
 Our results can be experimentally probed by a combination of rf-spectroscopy and three-body loss measurements of light impurities immersed in BECs. 
 The parameter regimes discussed in our work are experimentally feasible, if suitable magnetic fields are found that feature both very small interboson scattering length and a large boson-impurity scattering length. We expect our results to also hold, up to a quantitative shift, for larger interboson scattering lengths.
 
 Future interesting directions include the study of the real-time dynamics of the polaron \cite{drescher:2018,drescher:2020,drescher:2021} using Gaussian states, to understand how the Efimov cluster formation occurs in real-time and whether indeed resonant behaviour can be observed at the scattering length $a^{\ast}$. Furthermore, the scope of our results can be extended to include finite total momentum, explicit interboson repulsion and by going beyond the single-channel model. This will respectively allow the study of the dispersion relation of the polaron, the repulsive side of Feshbach resonances and closed-channel dominated resonances. Our methods can also be extended to rotating impurities \cite{schmidt:2015,schmidt:2016,lemeshko:2017} or bipolarons \cite{camacho:2018,naidon:2018,panochko:2021} to study the effect of interboson correlations in those systems.
 
Finally, it would be fascinating to explore further the connection to quenched BECs \cite{makotyn:2014,piatecki:2014,eismann:2016,klauss:2017,colussi:2018,eigen:2018,incao:2018,colussi:2020,musolino:2021}. In this work, we have shown that the formation of a polaron cloud around an impurity can lead to a modification of Efimov physics. One natural question to ask, is whether a similar effect occurs in BECs quenched to attractive scattering lengths. While one can certainly not employ the language of polarons or polaron clouds in this case, it is still pairwise correlations between the bosons that lead them to cluster closer together, which is also the essence of polaron physics. Hence, it will be interesting to explore whether a shift of Efimov resonances can also be observed in said scenarios. It may be possible to explore such effects by applying the cumulant expansion method as described in Ref.\ \cite{colussi:2020,musolino:2021} to attractive scattering lengths.
 
 Another question is how similar the polaronic instability is to the collapse of a BEC. In our model with bosonic interactions on the Bogoliubov level, the single impurity induces a first-order phase transition of the entire system at $a_{-}^{(min)}$. Even though this is prevented in practice by explicit interboson repulsion, it still indicates that the impurity can have a profound effect on the medium. Furthermore, the study of how large an Efimov cluster in a medium can become before it will lead to recombination and loss, remains another interesting open question. These questions highlight how the study of the impurity scenario can give new insights into the dynamics of quenched or collapsing BECs.
 
 \subsection*{Acknowledgements} The authors are thankful to Tommaso Guaita and Tao Shi for useful discussions. J.I.C. acknowledges funding from ERC Advanced Grant QENOCOBA under the EU Horizon
2020 program (Grant Agreement No. 742102). R.S. acknowledges support from the Deutsche Forschungsgemeinschaft (DFG,
German Research Foundation) under Germany’s Excellence Strategy–EXC–2111–390814868.

\bibliography{Paper.bib}

\onecolumngrid
\appendix

\section{Variable definitions} \label{app1}

In this appendix we provide definitions of the variables that are introduced by replacing the original bosons by the Bogoliubov quasiparticles in Eq.\ \ref{eq:Hamiltonian_Bog}.

The dispersion relation and operators are defined as:
\begin{align}
\omega_k &= \sqrt{\frac{k^4}{4m^2}+\frac{g_{B} n_0 k^2}{m}}, \\
\hat{a}_{\bm{k}}&=u_{k} \hat{b}_{\bm{k}} - v_{k}b^{\dagger}_{\bm{-k}},\\
u_k&=\sqrt{\frac{1}{2}(\frac{\omega_k+g_B n_0}{\omega_k}+1)}, \\
v_k&=\sqrt{\frac{1}{2}(\frac{\omega_k+g_B n_0}{\omega_k}-1)} .
\end{align}

The variables concerning the interaction of the impurity, with $W_k$,$V_{k,k'}^{(1)}$,$V_{k,k'}^{(2)}$ appearing in the Hamiltonian, and $a_0$ providing a mean-field shift of the scattering length due to the modified quasiparticle dispersion, are given by
\begin{align}
W_k &= \frac{k}{\sqrt{2m\omega_k}},\\
V_{k,k'}^{(1)}&= \frac{W_k  W_{k'}}{2} + \frac{1}{2 (W_k W_{k'})}, \\
V_{k,k'}^{(2)}&=  \frac{W_k  W_{k'}}{2} - \frac{1}{2 (W_k W_{k'})},\\
a_0^{-1}&=\frac{2 \Lambda}{\pi}-\frac{1}{\mu_r \pi}\int_0^{\Lambda} dk \frac{k^2 W_k^2}{\omega_k+\frac{k^2}{2M}}.
\end{align}

\section{The quartic Lee-Low-Pines term} \label{app2}

Here we explicitly give the definition of the quartic $\hat{\mathcal{H}}_{QLLP}$-term
which originates from the LLP-transformation on the impurity momentum operator in the spherical wave basis. It reads

\begin{multline} \label{eq:quarticterm}
\hat{\mathcal{H}}_{QLLP}=\int \int d\bm{k_1} d\bm{k_2} \frac{\bm{k}_1 \cdot \bm{k}_2}{2M}   \ \hat{b}^{\dagger}_{\bm{k_1}} \hat{b}^{\dagger}_{\bm{k_2}} \hat{b}_{\bm{k_1}} \hat{b}_{\bm{k_2}}= \\
\frac{1}{2M}\sum_{l_1 l_2 m_1 m_2} \int_0^{\Lambda} \int_0^{\Lambda} dk_1 dk_2  \frac{k_1 k_2}{\sqrt{(2l_1+1)(2l_1+3)(2l_2+1)(2l_2+3)}} \\
\Bigg[ \sqrt{(l_1-m_1+1)(l_1+m_1+1) (l_2-m_2+1)(l_2+m_2+1)} \\ [2 \hat{b}^{\dagger}_{k_1 (l_1+1) m_1} \hat{b}^{\dagger}_{k_2 l_2 m_2} \hat{b}_{k_1 l_1 m_1} \hat{b}_{k_2 (l_2+1)  m_2}  -  \hat{b}^{\dagger}_{k_1 (l_1+1) m_1} \hat{b}^{\dagger}_{k_2 (l_2+1) m_2} \hat{b}_{k_1 l_1 m_1} \hat{b}_{k_2 l_2  m_2}-  \hat{b}^{\dagger}_{k_1 l_1 m_1} \hat{b}^{\dagger}_{k_2 l_2 m_2} \hat{b}_{k_1 (l_1+1) m_1} \hat{b}_{k_2 (l_2+1)  m_2}]  \\
+ \sqrt{(l_1-m_1+1)(l_1-m_1+2) (l_2+m_2+1)(l_2+m_2+2)} \\ [\hat{b}^{\dagger}_{k_1 (l_1+1) (m_1-1)} 
\hat{b}^{\dagger}_{k_2 (l_2+1) (m_2+1)} \hat{b}_{k_1 l_1 m_1} \hat{b}_{k_2 l_2  m_2}+  \hat{b}^{\dagger}_{k_1 l_1 m_1} \hat{b}^{\dagger}_{k_2 l_2 m_2} \hat{b}_{k_1 (l_1+1) (m_1-1)} \hat{b}_{k_2 (l_2+1)  (m_2+1)}] 
 \\ + \sqrt{(l_1-m_1+1)(l_1-m_1+2) (l_2-m_2+1)(l_2-m_2+2)} \ \hat{b}^{\dagger}_{k_1 (l_1+1) (m_1-1)} \hat{b}^{\dagger}_{k_2 l_2 m_2} \hat{b}_{k_1 l_1 m_1} \hat{b}_{k_2 (l_2+1)  (m_2-1)}  
\\ +  \sqrt{(l_1+m_1+1)(l_1+m_1+2) (l_2+m_2+1)(l_2+m_2+2)} \ \hat{b}^{\dagger}_{k_1 (l_1+1) (m_1+1)} \hat{b}^{\dagger}_{k_2 l_2 m_2} \hat{b}_{k_1 l_1 m_1} \hat{b}_{k_2 (l_2+1)  (m_2+1)}   \Bigg].
\end{multline}

This expression only changes $l$ and $m$ by one at a time, coupling terms of $(l,m)$ terms to $(l \pm 1,m)$ or $(l \pm 1,m \pm 1)$. Furthermore, it respects angular momentum conservation. Together, this allows for an efficient numerical implementation.

\section{Relevant variables for the time-evolution of Gaussian States}

Here we present the relevant variables needed to perform the time evolution of the Gaussian State in case of $n_0=0$. We find  that even though the $\hat{\mathcal{H}}_{QLLP}$-term defined in App.\ \ref{app2} is complicated, it simplifies due to the spherical symmetry of the problem:

\begin{equation}
\eta_{k}=\frac{k^2}{2\mu_r} \phi_{k}+
\frac{gk}{2 \pi^2} \int_0^{\Lambda}dk'  k' \phi_{k'}+\frac{k}{2M}\int_0^{\Lambda} k' 2( G_{kk'(l=1)} \phi_{k'}-  F_{kk'(l=1)} \phi_{k'}^{\ast}),
\end{equation}

\begin{align}
\Eps_{k_1k_2l}&=\delta_{k_1k_2}\frac{k_1^2}{2 \mu_r}+ \delta_{l,0} \frac{gk_1k_2}{2 \pi^2} + \frac{(l+1)k_i k_j}{(2l+1)M}  G_{k_1k_2(l+1)}+  \frac{l k_1 k_2}{(2l+1)M} [\delta_{l,1}\phi_{k_1}\phi_{k_2}^{\ast}+ G_{k_1k_2(l-1)}],  \\
\Delta_{k_1k_2l}&=-\frac{(l+1)k_1 k_2}{(2l+1)M} F_{k_1k_2(l+1)}- \frac{l k_1 k_2}{(2l+1)M} [\delta_{l,1} \phi_{k_1} \phi_{k_2}+ F_{k_1 k_2 (l-1)} ].
\end{align}

\end{document}